\documentclass[10pt]{article}
\usepackage{graphicx}
\usepackage{amsmath}
\usepackage{amssymb}
\usepackage{caption2}
\begin{document}
\begin{center}
{\large {\bf \sc{The fully-light vector tetraquark states with explicit P-wave via the QCD sum rules}}} \\[2mm]
Qi Xin${}^{*\dagger}$, Zhi-Gang Wang${}^*$\footnote{E-mail: zgwang@aliyun.com. },

Department of Physics, North China Electric Power University, Baoding 071003, P. R. China${}^*$\\
School of Nuclear Science and Engineering, North China Electric Power University, Beijing 102206, P. R. China${}^\dagger$
\end{center}

\begin{abstract}
In this paper, we apply the QCD sum rules to study the vector fully-light tetraquark states with an
explicit P-wave between the diquark and antidiquark pair. We observed that the  $C\gamma_\alpha\otimes\stackrel{\leftrightarrow}{\partial}_\mu\otimes\gamma^\alpha C$ (or $C\gamma_\alpha\otimes\stackrel{\leftrightarrow}D_\mu\otimes\gamma^\alpha C$) type current with fully-strange quarks couples potentially to a tetraquark state with the mass $2.16 \pm 0.14 \,\rm{GeV}$, which supports  assigning the $Y(2175)/\phi(2170)$ as the diquark-antidiquark type tetraquark state with the $J^{PC}=1^{--}$. The $qs\bar{q}\bar{s}$ and $ss\bar{s}\bar{s}$ vector tetraquark states with the structure $C\gamma_\mu\otimes \stackrel{\leftrightarrow}{\partial}_\alpha \otimes\gamma^\alpha C + C\gamma^\alpha \otimes\stackrel{\leftrightarrow}{\partial}_\alpha \otimes\gamma_\mu$ (or $C\gamma_\mu\otimes \stackrel{\leftrightarrow}D_\alpha \otimes\gamma^\alpha C + C\gamma^\alpha \otimes\stackrel{\leftrightarrow}D_\alpha \otimes\gamma_\mu$) are consistent with the $X(2200)$ and $X(2400)$ respectively, which lie in the  region $2.20$ to $2.40\,\rm{GeV}$. The central values of the masses of the fully-strange vector  tetraquark states with an explicit P-wave are about  $2.16-3.13\,\rm{GeV}$ (or $2.16-3.16\,\rm{GeV}$), and the predictions for other fully-light vector tetraquark states with and without hidden-strange are also presented.
\end{abstract}



\section{Introduction}
In 2006, the BaBar collaboration studied the cross section $e^+ e^- \to \phi f_{0}(980)$ and discovered a structure near threshold compatible with the quantum numbers \ $J^{PC}=1^{--}$ for the first time, the mass is $M = 2.175 \pm 0.010\pm 0.015\,\rm{GeV}$ and the decay width is $\Gamma = 58\pm 16\pm 20\,\rm{MeV}$ \cite{BabarY2175-2006}, thereafter, it was named as the $Y(2175)$ \cite{ssss-ZGWANG-2006}. It is a suitable fully-light tetraquark  candidate \cite{ssss-ZGWANG-2006} and the experimental results have attracted extensive interest as there also exist other possible interpretations beyond the tetraquark state \cite{Review-CPC}. Later, the BESII collaboration  confirmed the presence of the $Y(2175)$ by providing the Breit-Wigner  mass and width:  $M  =  2.186\pm0.010 \pm0.006\,\rm{GeV}$ and $\Gamma =  65\pm23 \pm17\,\rm{MeV}$ \cite{BESIIY2175-2007}. And the Belle collaboration also conformed the $Y(2175)$, in addition, there is a cluster of events near $2.4\,\rm{GeV}$ \cite{BelleY2175-2008}. Shen and Yuan studied  the combined data from the Belle and BaBar collaborations and observed an evidence for the structure $X(2400)$ with a mass $  2.436\pm0.026 \,\rm{GeV}$ and a width $  121\pm35 \,\rm{MeV}$ \cite{Y2400-2008-combine}.

In 2019, the BESIII collaboration measured the  cross section of the process $e^+ e^- \to K^+K^-$ ($K^+K^-K^+K^-$) and  observed a resonance structure, $X(2240)$, with the Breit-Wigner  mass and width: $M  = 2239.2\pm7.1\pm 11.3 \,\rm{MeV}$ ($2.232\,\rm{GeV}$) and $\Gamma =  139.8\pm12.3 \pm20.6 \,\rm{MeV}$ \cite{BESIIIY2175(2239)-2018} (\cite{BESIII2232-2019}). The resonant parameters   differ from the world average
parameters of the $\rho(2150)$ and $\phi(2170)$ by more than $3\sigma$ in
mass and more than $2\sigma$ in width \cite{BESIIIY2175(2239)-2018}. Furthermore, the isovector resonance
$\rho(2150)$ is not well established \cite{PDG-2020}.
In 2020, the BaBar collaboration measured the same cross section in conjunction with previous BaBar results for the relevant  processes,   and obtained similar Breit-Wigner mass and width for the isovector resonance $\rho(2230)$:  $M  =  2232\pm8\pm 9 \,\rm{MeV}$ and $\Gamma =  133\pm14 \pm4 \,\rm{MeV}$  \cite{BABAR2232-2020}.
Recently, the BESIII collaboration measured the cross section $e^+ e^- \to \Lambda\bar{\Lambda} \eta$, and observed a resonance structure, $X(2400)$ with the $J^{PC}=1^{--}$ , with the Breit-Wigner  mass and width: $M=2356\pm7\pm 17 \,\rm{MeV}$ and  $\Gamma=304 \pm  28 \pm 54\,\rm{MeV}$, respectively \cite{BESIII2356-2022}.

Up to now, an increasing number of potential fully-light exotic hadron candidates  have been discovered. The $X(1835)$ previously observed by the BESII collaboration \cite{BES-1835} was verified with the statistical significance more than $20\sigma$ \cite{BESIIIX1835/2120/2370-2010}. Additionally, the $X(2120)$ and  $X(2370)$ were observed in the $\pi^+ \pi^- \eta'$ invariant mass spectrum   in the decays  $J/\psi \to \gamma \pi^+ \pi^- \eta'$ \cite{BESIIIX1835/2120/2370-2010}. In 2016, the pseudoscalar states $\eta(2100)$ and $X(2500)$  were detected, the scalar state $f_0(2100)$ and tensor states $f_2(2010)$, $f_2(2300)$ and $f_2(2340)$ were  also detected in the process $J/\psi\to \gamma \phi \phi$ \cite{BESIIIX2100/2500etc-2016}. As the pieces of evidences  began to accumulate, the theoretical physicists  are obliged to investigate the fully-light tetraquark states.
In short, there have been several possible theoretical interpretations for the $Y(2175)/\phi(2170)$, such as the tetraquark state \cite{ssss-ZGWANG-2006,Y2175CHX2008,Y2175CHX2018,Y2175CHX2022,X2100/2239-AZIZI-2019,
Y2175-AZIZI-2019,ssss-LQF-2019,2170-LXQ-2018,2175-1536-PJL-2010}, $\Lambda\bar{\Lambda}$ baryonium \cite{2175-PJL-2013,2170-LQF-2017}, conventional $s\bar{s}$ state \cite{2175-DGJ-2007-1,phi-CQP-2019}, $\phi K \bar K$ resonance state \cite{phiKK-2008,phiKK-2009} and strangeonium hybrid \cite{2175-DGJ-2007}.

In previous works,  the QCD sum rules have been used to explore the tetraquark states
$ss\bar{s}\bar{s}$ with the quantum numbers $J^P=0^{++}$, $1^{--}$, $1^{+-}$, $2^{++}$, et al  \cite{ssss-ZGWANG-2006,Y2175CHX2008,Y2175CHX2018,Y2175CHX2022,X2100/2239-AZIZI-2019,ssss-ZGWANG-2019}.
In Refs.\cite{ssss-ZGWANG-2006,Y2175CHX2008,Y2175CHX2018,X2100/2239-AZIZI-2019},
the $Y(2175)/\phi(2170)$  is interpreted as the $ss\bar{s}\bar{s}$  vector tetraquark state (or as the $sq\bar{s}\bar{q}$ state \cite{Y2175-AZIZI-2019}) with an implicit P-wave (or with an explicit P-wave in the (anti)diquark constituent \cite{Y2175CHX2022}) according to the calculations via the QCD sum rules. On the other hand, the mass spectrum and strong decays of the $ss\bar{s}\bar{s}$ states with the $S$-wave and $P$-wave are studied in the relativized quark model, the assignment of the $Y(2175)/\phi(2170)$ as a $ss\bar{s}\bar{s}$ candidate cannot be excluded \cite{ssss-LQF-2019,2170-LXQ-2018}. In the flux-tube model, the ${}^3P_0$ model, the modified GI quark model and quark pair creation (QPC) model, the state $Y(2175)/\phi(2170)$ is assigned as the hidden strange $ss\bar{s}\bar{s}$ tetraquark state \cite{2175-1536-PJL-2010}, the $\Lambda\bar{\Lambda}$ baryonium \cite{2175-PJL-2013,2170-LQF-2017}, and the $2^3D_1s\bar{s}$ meson \cite{2175-DGJ-2007-1,phi-CQP-2019}, respectively. For more details and literatures, one can consult the review \cite{Review-CPC}.

 The diquarks $\varepsilon^{ijk}q^{T}_j C\Gamma q^{\prime}_k$ in the color antitriplet have five structures in Dirac spinor space, $C\Gamma=C\gamma_5$, $C$, $C\gamma_\mu \gamma_5$, $C\gamma_\mu$ and $C\sigma_{\mu\nu}$ for the scalar, pseudoscalar, vector, axialvector and tensor diquarks, respectively, where the $i$, $j$ and $k$ are color indexes. The favored or stable configurations are the scalar $\varepsilon^{ijk}q^{T}_j C\gamma_5 q^{\prime}_k$  and axialvector $\varepsilon^{ijk}q^{T}_j C\gamma_\mu q^{\prime}_k$ diquark states based on the QCD sum rules \cite{WZG-CTP-SA}, for the diquarks containing the same flavor, $\varepsilon^{ijk}q^{T}_j C\gamma_5 q_k = 0$, we give priority to the axialvector diquarks $\varepsilon^{ijk}q^{T}_j C\gamma_{\mu} q_k$ as the basic constituents. However, the diquarks $q^{T}_j C\gamma_5 q_k $ in the color sextet can exist indeed
 due to Fermi-Dirac statistics, although the attractive (repulsive) force from the one-gluon exchange  favors (disfavors) formation of the diquarks in the color antitriplet (sextet) \cite{Color-Spin,Jaffe1977-1,Jaffe1977-2}.

 In case of the negative-parity is concerned, the relative P-waves in the diquarks can be implied in the underlined $\gamma_5$ in the $C\gamma_5 \underline{\gamma_5} \otimes \gamma_\mu C$-type and $C\gamma_5 \otimes \underline{\gamma_5}\gamma_\mu C$-type currents or in the underlined  $\gamma^\alpha$ in the $C\gamma_\alpha \underline{\gamma^\alpha} \otimes \gamma_\mu C$-type currents \cite{Y4260-ZGWANG-2018}, or  in the vector components in the $C\sigma_{\mu\nu}$ and $C\sigma_{\mu\nu}\gamma_5$ type diquarks \cite{WZG-CTP-4100}.
 In Refs.\cite{Y2175CHX2008,Y2175CHX2018}, Chen et al
  take  the diquarks both in the  color antitriplet and sextet as the basic constituents to construct two vector currents to interpolate the $Y(2175)$, where the P-wave is implied in the diquarks, then take account of the mixing effects between the two currents, and obtain the lowest masses $2.41 \pm 0.25 \,\rm{GeV}$ and $2.34 \pm 0.17\,\rm{GeV}$ for the vector $ss\bar{s}\bar{s}$ tetraquark states,  the central values are larger than  the experimental data. In Ref.\cite{ssss-ZGWANG-2019}, we choose the $C\gamma_\mu\otimes \gamma_\nu C-C\gamma_\nu\otimes \gamma_\mu C$-type current without introducing an explicit or implicit P-wave to interpolate the $Y(2175)$, the P-wave is only implicitly embodied in the non-vanishing couplings to the vector tetraquark states.  The calculations also lead to larger mass for the
  vector $ss\bar{s}\bar{s}$ tetraquark configuration having an implicit P-wave compared  to the experimental  mass of the $Y(2175)/\phi(2170)$. In Ref.\cite{ssss-ZGWANG-2006}, we choose the $\gamma_\mu \otimes \underline{\gamma_5} \gamma_5$-type current in the color octet-octet to interpolate the $Y(2175)$, where
   the P-wave is implied in the underlined $\gamma_5$, and reproduce the experimental mass of the $Y(2175)$, however, the pole contribution is not large enough.

  So we propose to introduce the explicit  P-wave to study the fully-light vector tetraquark states, and construct
   the $\stackrel{\leftrightarrow}{\partial}_\mu C\gamma_5\otimes \gamma_5C$-type, $C\gamma_5\otimes\stackrel{\leftrightarrow}{\partial}_\mu \gamma_5C$-type,  $\stackrel{\leftrightarrow}{\partial}_\mu C\gamma_\alpha \otimes \gamma^\alpha C$-type and  $C\gamma_\alpha \otimes \stackrel{\leftrightarrow}{\partial}_\mu \gamma^\alpha C$-type four-quark vector currents, where $\stackrel{\leftrightarrow}{\partial}_\mu=\stackrel{\rightarrow}{\partial}_\mu-\stackrel{\leftarrow}{\partial}_\mu $ \cite{Y4260P-wave-ZGWANG-2018,Y4220P-wave-ZGWANG-2018}. The additional P-wave in the non-relativistic quark model can alter the parity by adding a factor $(-)^L=-$, where $L=1$ is the angular momentum.

By comparing Ref.\cite{Y4220P-wave-ZGWANG-2018} with Ref.\cite{NPB-ZGWANG-2021},
we reveal that the masses of the $C \otimes \gamma_\mu C \pm C \gamma_\mu \otimes  C$-type,
$C\gamma_5 \otimes \gamma_5\gamma_\mu C \pm C \gamma_\mu\gamma_5 \otimes  \gamma_5 C$-type
and
$C\gamma_\mu \otimes \gamma_\nu C - C \gamma_\nu \otimes  \gamma_\mu C$-type vector hidden-charm tetraquark states are quite different from the masses of the $C \gamma_5 \stackrel{\leftrightarrow}{\partial}_\mu  \gamma_5 C$-type,
$C \gamma_\alpha \otimes\stackrel{\leftrightarrow}{\partial}_\mu\otimes \gamma^\alpha C$-type,
$C \gamma_\mu \otimes\stackrel{\leftrightarrow}{\partial}_\alpha\otimes \gamma^\alpha C
+ C \gamma^\alpha \otimes\stackrel{\leftrightarrow}{\partial}_\alpha\otimes \gamma_\mu C$-type and
$C \gamma_5 \otimes\stackrel{\leftrightarrow}{\partial}_\mu\otimes \gamma_\nu C
+ C \gamma_\nu \stackrel{\leftrightarrow}{\partial}_\mu \gamma_5 C
-C \gamma_5 \otimes\stackrel{\leftrightarrow}{\partial}_\nu\otimes \gamma_\mu C
- C \gamma_\mu \otimes\stackrel{\leftrightarrow}{\partial}_\nu\otimes \gamma_5 C$-type vector hidden-charm tetraquark states. The $Y(2175)/\phi(2170)$ could be viewed as a $s\bar s$ analogue of the $Y(4260)$, or as a $s\bar ss\bar s$ state decays primarily to the $\phi f_0(980)$.

Now we extend our previous work on the $Y(4220/4260)$,  $Y(4320/4360)$, $Y(4390)$ and $Y(4660)$ carried out  in Ref.\cite{Y4220P-wave-ZGWANG-2018},
where we study  the hidden-charm vector tetraquark states with an explicit P-wave between the diquark and
antidiquark using the QCD sum rules in a systematic way, to explore the fully-light tetraquark states. More explicitly,
we extend the doubly-heavy and doubly-light quarks to the case where the four valence  quarks are all light quarks, and study the $ss\bar{s}\bar{s}$, $sq\bar{s}\bar{q}$,
and $qq\bar{q}\bar{q}$ tetraquark states with an explicit P-wave via the QCD sum rules,  then compare the predictions  to
the  corresponding ones  without an explicit P-wave \cite{Y2175CHX2008,Y2175CHX2018,ssss-ZGWANG-2019}.

Generally speaking, we can choose either the partial  derivative or covariant derivative to construct  the interpolating currents. The currents with  covariant derivative $D_\mu$ are gauge invariant, but does not favor   interpreting  the $\stackrel{\leftrightarrow}{D}_\mu=\stackrel{\rightarrow}{\partial}_\mu-ig_sG_\mu-\stackrel{\leftarrow}{\partial}_\mu-ig_sG_\mu $ as  angular momentum.
 The currents with partial derivative $\partial_\mu$ are not gauge covariant, but favors  interpreting  the $\stackrel{\leftrightarrow}{\partial}_\mu=\stackrel{\rightarrow}{\partial}_\mu-\stackrel{\leftarrow}{\partial}_\mu$ as  angular momentum, furthermore, the covariant derivative $D_\mu$ leads to some hybrid components in the hadron states due to the gluon field $G_\mu$.
In this work, we present the results with both the partial  derivatives  $\partial_\mu$ and covariant derivatives  $D_\mu$ for completeness.

The paper is organized as follows: in section 2, we derive the QCD sum rules for the vector tetraquark states with explicit P-waves; in section 3, we show the numerical results and discussions; and in section 4, we draw conclusions.

\section{QCD sum rules for the fully-light vector tetraquark states}

In the following, we write down the two-point correlation functions $\Pi_{\mu\nu}(p)$
and $\Pi_{\mu\nu\alpha\beta}(p)$ in the QCD sum rules,
\begin{eqnarray}
\Pi_{\mu\nu}(p)&=&i\int d^4x e^{ip \cdot x} \langle0|T\left\{J_\mu(x)J_\nu^{\dagger}(0)\right
\}|0\rangle \, , \\
\Pi_{\mu\nu\alpha\beta}(p)&=&i\int d^4x e^{ip \cdot x} \langle0|T\left\{J_{\mu\nu}(x)
J_{\alpha\beta}^{\dagger}(0)\right\}|0\rangle \, ,
\end{eqnarray}
where $J_\mu(x)=J^{1,2,3,4}_{\mu,ss\bar{s}\bar{s}/qs\bar{q}\bar{s}/qq\bar{q}\bar{q}}(x)$,
$J^5_{\mu,qs\bar{q}\bar{s}}(x)$, $J_{\mu\nu}(x)=J^6_{\mu\nu,qs\bar{q}\bar{s}}(x)$,
\begin{eqnarray}\label{J-ssss}
J^1_{\mu,ss\bar{s}\bar{s}}(x)&=&\varepsilon^{ijk}\varepsilon^{imn}s^{Tj}(x)C
\gamma_\alpha s^k(x)\stackrel{\leftrightarrow}{\partial}_\mu \bar{s}^m(x)\gamma^\alpha C
\bar{s}^{Tn}(x) \, ,\nonumber\\
J^2_{\mu,ss\bar{s}\bar{s}}(x)&=&\frac{\varepsilon^{ijk}\varepsilon^{imn}}{\sqrt{2}}
\left[\begin{array}{l} \quad s^{Tj}(x)C\gamma_\mu s^k(x)\stackrel{\leftrightarrow}{\partial}_\alpha
\bar{s}^m(x)\gamma^\alpha C \bar{s}^{Tn}(x)  \\
+\, s^{Tj}(x)C\gamma^\alpha s^k(x)\stackrel{\leftrightarrow}{\partial}_\alpha \bar{s}^m(x)\gamma_\mu C
\bar{s}^{Tn}(x) \end{array} \right]\, ,\nonumber\\
J^3_{\mu,ss\bar{s}\bar{s}}(x)&=&\varepsilon^{ijk}\varepsilon^{imn}s^{Tj}(x)C\sigma_{\alpha\beta}
s^k(x)\stackrel{\leftrightarrow}{\partial}_\mu \bar{s}^m(x) \sigma^{\alpha\beta} C \bar{s}^{Tn}(x) \, ,\nonumber\\
J^4_{\mu,ss\bar{s}\bar{s}}(x)&=&\frac{\varepsilon^{ijk}\varepsilon^{imn}}{\sqrt{2}}\left[\begin{array}{l} \quad
s^{Tj}(x)C\sigma_{\mu\nu} s^k(x)\stackrel{\leftrightarrow}{\partial}_\alpha \bar{s}^m(x)\sigma^{\alpha\nu} C
\bar{s}^{Tn}(x)  \\
+\, s^{Tj}(x)C\sigma^{\alpha\nu} s^k(x)\stackrel{\leftrightarrow}{\partial}_\alpha \bar{s}^m(x)\sigma_{\mu\nu} C
\bar{s}^{Tn}(x)\end{array} \right]\, ,
\end{eqnarray}

\begin{eqnarray}
J^1_{\mu,qs\bar{q}\bar{s}}(x)&=&\varepsilon^{ijk}\varepsilon^{imn}q^{Tj}(x)C
\gamma_\alpha s^k(x)\stackrel{\leftrightarrow}{\partial}_\mu \bar{q}^m(x)
\gamma^\alpha C \bar{s}^{Tn}(x) \, ,\nonumber\\
J^2_{\mu,qs\bar{q}\bar{s}}(x)&=&\frac{\varepsilon^{ijk}\varepsilon^{imn}}
{\sqrt{2}}\left[\begin{array}{l} \quad q^{Tj}(x)C\gamma_\mu s^k(x)\stackrel{\leftrightarrow}{\partial}_\alpha
\bar{q}^m(x)\gamma^\alpha C \bar{s}^{Tn}(x) \\
 + \,q^{Tj}(x)C\gamma^\alpha s^k(x)\stackrel{\leftrightarrow}{\partial}_\alpha
 \bar{q}^m(x)\gamma_\mu C \bar{s}^{Tn}(x)\end{array} \right]\, ,\nonumber\\
J^3_{\mu,qs\bar{q}\bar{s}}(x)&=&\varepsilon^{ijk}\varepsilon^{imn}q^{Tj}(x)C
\sigma_{\alpha\beta} s^k(x)\stackrel{\leftrightarrow}{\partial}_\mu \bar{q}^m(x)
\sigma^{\alpha\beta} C \bar{s}^{Tn}(x) \, ,\nonumber\\
J^4_{\mu,qs\bar{q}\bar{s}}(x)&=&\frac{\varepsilon^{ijk}\varepsilon^{imn}}{\sqrt{2}}\left[\begin{array}{l}
\quad q^{Tj}(x)C\sigma_{\mu\nu} s^k(x)\stackrel{\leftrightarrow}{\partial}_\alpha
\bar{q}^m(x)\sigma^{\alpha\nu} C \bar{s}^{Tn}(x) \\
 + \, q^{Tj}(x)C\sigma^{\alpha\nu} s^k(x)\stackrel{\leftrightarrow}{\partial}_\alpha \bar{q}^m(x)\sigma_{\mu\nu} C
 \bar{s}^{Tn}(x)\end{array}\right]\, ,\nonumber\\
J^5_{\mu,qs\bar{q}\bar{s}}(x)&=&\varepsilon^{ijk}\varepsilon^{imn}q^{Tj}(x)C\gamma_5
s^k(x)\stackrel{\leftrightarrow}{\partial}_\mu \bar{q}^m(x)\gamma_5 C \bar{s}^{Tn}(x) \, ,\nonumber\\
J^6_{\mu\nu,qs\bar{q}\bar{s}}(x)&=&\frac{\varepsilon^{ijk}\varepsilon^{imn}}{2}\left[\begin{array}{l} \quad
q^{Tj}(x)C\gamma_5 s^k(x)\stackrel{\leftrightarrow}{\partial}_\mu \bar{q}^m(x)\gamma_\nu C \bar{s}^{Tn}(x)  \\
+ \,q^{Tj}(x)C\gamma_\nu s^k(x)\stackrel{\leftrightarrow}{\partial}_\mu \bar{q}^m(x)\gamma_5 C \bar{s}^{Tn}(x)   \\
-\, q^{Tj}(x)C\gamma_5 s^k(x)\stackrel{\leftrightarrow}{\partial}_\nu \bar{q}^m(x)\gamma_\mu C \bar{s}^{Tn}(x)  \\
-\, q^{Tj}(x)C\gamma_\mu s^k(x)\stackrel{\leftrightarrow}{\partial}_\nu \bar{q}^m(x)\gamma_5 C \bar{s}^{Tn}(x)\end{array} \right]\, ,
\end{eqnarray}

\begin{eqnarray}\label{J-qqqq}
J^1_{\mu,qq\bar{q}\bar{q}}(x)&=&\varepsilon^{ijk}\varepsilon^{imn}q^{Tj}(x)C\gamma_\alpha
q^k(x)\stackrel{\leftrightarrow}{\partial}_\mu \bar{q}^m(x)\gamma^\alpha C \bar{q}^{Tn}(x) \, ,\nonumber\\
J^2_{\mu,qq\bar{q}\bar{q}}(x)&=&\frac{\varepsilon^{ijk}\varepsilon^{imn}}{\sqrt{2}}\left[\begin{array}{l} \quad
q^{Tj}(x)C\gamma_\mu q^k(x)\stackrel{\leftrightarrow}{\partial}_\alpha \bar{q}^m(x)\gamma^\alpha C \bar{q}^{Tn}(x) \\
+ \,q^{Tj}(x)C\gamma^\alpha q^k(x)\stackrel{\leftrightarrow}{\partial}_\alpha \bar{q}^m(x)\gamma_\mu C
\bar{q}^{Tn}(x)\end{array} \right]\, ,\nonumber\\
J^3_{\mu,qq\bar{q}\bar{q}}(x)&=&\varepsilon^{ijk}\varepsilon^{imn}q^{Tj}(x)C\sigma_{\alpha\beta}
q^k(x)\stackrel{\leftrightarrow}{\partial}_\mu \bar{q}^m(x) \sigma^{\alpha\beta} C \bar{q}^{Tn}(x) \, ,\nonumber\\
J^4_{\mu,qq\bar{q}\bar{q}}(x)&=&\frac{\varepsilon^{ijk}\varepsilon^{imn}}{\sqrt{2}}\left[\begin{array}{l}
\quad q^{Tj}(x)C\sigma_{\mu\nu} s^k(x)\stackrel{\leftrightarrow}{\partial}_\alpha \bar{q}^m(x)\sigma^{\alpha\nu} C
\bar{q}^{Tn}(x)  \\
+ \, q^{Tj}(x)C\sigma^{\alpha\nu} q^k(x)\stackrel{\leftrightarrow}{\partial}_\alpha \bar{q}^m(x)\sigma_{\mu\nu} C
\bar{q}^{Tn}(x)\end{array} \right]\, ,
\end{eqnarray}
where the $i$, $j$, $k$, $m$, $n$ are color indexes,  the $C$ is the charge conjugation matrix, and
$\stackrel{\leftrightarrow}{\partial}_\mu=\stackrel{\rightarrow}{\partial}_\mu-\stackrel{\leftarrow}{\partial}_\mu $. We choose the tensor diquark operators $\varepsilon^{ijk}s^{Tj}(x)C\sigma_{\mu\nu} s^k(x)$, $\varepsilon^{ijk}q^{Tj}(x)C\sigma_{\mu\nu} s^k(x)$ and $\varepsilon^{ijk}q^{Tj}(x)C\sigma_{\mu\nu} q^k(x)$ beyond the axialvector diquark operators to construct the four-quark currents, as they also exist due to Fermi-Dirac statistics.

We can take a simple replacement  $\stackrel{\leftrightarrow}{\partial}_\mu \to \stackrel{\leftrightarrow}{D}_\mu$ in Eqs.\eqref{J-ssss}-\eqref{J-qqqq} to acquire the corresponding gauge invariant currents,
\begin{eqnarray}
J_\mu(x)&\to& J_\mu(x)\mid_{\stackrel{\leftrightarrow}{\partial} \to \stackrel{\leftrightarrow}{D}}\, ,\nonumber\\
J_{\mu\nu}(x)&\to& J_{\mu\nu}(x)\mid_{\stackrel{\leftrightarrow}{\partial} \to \stackrel{\leftrightarrow}{D}}\, ,
\end{eqnarray}
just like what we have done in Refs.\cite{WZG-Dwave-baryon,XQWZG-Pwave-Bbaryon}. In this work, we choose the currents with the partial derivatives and covariant derivatives respectively to interpolate the vector tetraquark states.

    Under parity transform $\widehat{P}$ and charge conjugation transform $\widehat{C}$, the currents $J_\mu(x)$ and $J_{\mu\nu}(x)$ have the properties,

\begin{eqnarray}\label{J-parity}
\widehat{P} J_\mu(x)\widehat{P}^{-1}&=&+J^\mu(\tilde{x}) \, , \nonumber\\
\widehat{P} J_{\mu\nu}(x)\widehat{P}^{-1}&=&-J^{\mu\nu}(\tilde{x}) \, ,
\end{eqnarray}
\begin{eqnarray}
\widehat{C}J_{\mu}(x)\widehat{C}^{-1}&=&- J_{\mu}(x) \, , \nonumber\\
\widehat{C}J_{\mu\nu}(x)\widehat{C}^{-1}&=&- J_{\mu\nu}(x) \, ,
\end{eqnarray}
where  $x^\mu=(t,\vec{x})$ and $\tilde{x}^\mu=(t,-\vec{x})$.

On the hadron side, we acquire the hadronic representation by inserting a complete set of intermediate hadronic states with the same quantum numbers as the current operators $J_\mu(x)$ and $J_{\mu\nu}(x)$ into the correlation functions $\Pi_{\mu\nu}(p)$ and $\Pi_{\mu\nu\alpha\beta}(p)$ \cite{QCDSR-SVZ79,QCDSR-Reinders85}, and reach the following results by separating the ground state contributions of the vector tetraquark states,

\begin{eqnarray}
\Pi_{\mu\nu}(p)&=&\frac{\lambda_{Y}^2}{M_{Y}^2-p^2}\left(-g_{\mu\nu} +\frac{p_\mu p_\nu}{p^2}\right) + \cdots \, \, ,\nonumber\\
&=&\Pi_Y(p^2)\left(-g_{\mu\nu} +\frac{p_\mu p_\nu}{p^2}\right) +\cdots  \, , \\
\Pi_{\mu\nu\alpha\beta}(p)&=&\frac{\lambda_{ Y}^2}{M_{Y}^2\left(M_{Y}^2-p^2\right)}\left(p^2g_{\mu\alpha}g_{\nu\beta} -p^2g_{\mu\beta}g_{\nu\alpha} -g_{\mu\alpha}p_{\nu}p_{\beta}-g_{\nu\beta}p_{\mu}p_{\alpha}+g_{\mu\beta}p_{\nu}p_{\alpha}+g_{\nu\alpha}p_{\mu}p_{\beta}\right) \nonumber\\
&&+\frac{\lambda_{ Z}^2}{M_{Z}^2\left(M_{Z}^2-p^2\right)}\left( -g_{\mu\alpha}p_{\nu}p_{\beta}-g_{\nu\beta}p_{\mu}p_{\alpha}+g_{\mu\beta}p_{\nu}p_{\alpha}+g_{\nu\alpha}p_{\mu}p_{\beta}\right) +\cdots \, , \nonumber\\
&=&\widetilde{\Pi}_{Y}(p^2)\left(p^2g_{\mu\alpha}g_{\nu\beta} -p^2g_{\mu\beta}g_{\nu\alpha} -g_{\mu\alpha}p_{\nu}p_{\beta}-g_{\nu\beta}p_{\mu}p_{\alpha}+g_{\mu\beta}p_{\nu}p_{\alpha}+g_{\nu\alpha}p_{\mu}p_{\beta}\right) \nonumber\\
&&+\widetilde{\Pi}_{Z}(p^2)\left( -g_{\mu\alpha}p_{\nu}p_{\beta}-g_{\nu\beta}p_{\mu}p_{\alpha}+g_{\mu\beta}p_{\nu}p_{\alpha}+g_{\nu\alpha}p_{\mu}p_{\beta}\right) \, ,
\end{eqnarray}
where the pole residues  $\lambda_{Y}$ and $\lambda_{Z}$ are  defined by
\begin{eqnarray}
\langle 0|J_\mu(0)|Y(p)\rangle &=&\lambda_{Y} \,\varepsilon_\mu \, , \nonumber\\
  \langle 0|J_{\mu\nu}(0)|Y(p)\rangle &=& \frac{\lambda_{Y}}{M_{Y}} \, \varepsilon_{\mu\nu\alpha\beta} \, \varepsilon^{\alpha}p^{\beta}\, , \nonumber\\
 \langle 0|J_{\mu\nu}(0)|Z(p)\rangle &=& \frac{\lambda_{Z}}{M_{Z}} \left(\varepsilon_{\mu}p_{\nu}-\varepsilon_{\nu}p_{\mu} \right)\, ,
\end{eqnarray}
the $\varepsilon_\mu$ are the polarization vectors  of the  vector and axialvector tetraquark states $Y$ and $Z$ with the $J^{PC}=1^{--}$ and $1^{+-}$, respectively.
Now we project out the components $\Pi_{Y}(p^2)$ and $\Pi_{Z}(p^2)$ by introducing the operators $P_{Y}^{\mu\nu\alpha\beta}$ and $P_{Z}^{\mu\nu\alpha\beta}$,
\begin{eqnarray}
\Pi_{Y}(p^2)&=&p^2\widetilde{\Pi}_{Y}(p^2)=P_{Y}^{\mu\nu\alpha\beta}\Pi_{\mu\nu\alpha\beta}(p) \, , \nonumber\\
\Pi_{Z}(p^2)&=&p^2\widetilde{\Pi}_{Z}(p^2)=P_{Z}^{\mu\nu\alpha\beta}\Pi_{\mu\nu\alpha\beta}(p) \, ,
\end{eqnarray}
where
\begin{eqnarray}
P_{Y}^{\mu\nu\alpha\beta}&=&\frac{1}{6}\left( g^{\mu\alpha}-\frac{p^\mu p^\alpha}{p^2}\right)\left( g^{\nu\beta}-\frac{p^\nu p^\beta}{p^2}\right)\, , \nonumber\\
P_{Z}^{\mu\nu\alpha\beta}&=&\frac{1}{6}\left( g^{\mu\alpha}-\frac{p^\mu p^\alpha}{p^2}\right)\left( g^{\nu\beta}-\frac{p^\nu p^\beta}{p^2}\right)-\frac{1}{6}g^{\mu\alpha}g^{\nu\beta}\, .
\end{eqnarray}
In this paper, we choose the components $\Pi_{Y}(p^2)$ to study the fully-light tetraquark states with the $J^{PC}=1^{--}$, and acquire the hadronic spectral representation through dispersion relation,
\begin{eqnarray}
\Pi_{Y}(p^2) &=& \frac{1}{\pi} \int_0^\infty \frac{{\rm Im}\Pi_{Y}(s)}{s-p^2}\, .
\end{eqnarray}

On the QCD side, if we take  the current $J_\mu(x)=J^1_{\mu,ss\bar{s}\bar{s}}(x)$ with the partial derivatives as an example, the correlation function $\Pi_{\mu\nu}(p)$   can be written as
\begin{eqnarray}\label{example}
\Pi_{\mu\nu}(p)&=&-4i\varepsilon^{ijk}\varepsilon^{imn}\varepsilon^{i^{\prime}j^{\prime}k^{\prime}}\varepsilon^{i^{\prime}m^{\prime}n^{\prime}}\int d^4x\, e^{ip \cdot x} \nonumber\\
&&\left[\begin{array}{l}+{\rm Tr}\! \left\{\gamma_\alpha S^{kk^{\prime}}(x)\gamma_\beta CS^{jj^{\prime}T}(x)C \right\}  \partial_\mu \partial_\nu {\rm Tr}\!\left\{\gamma^\beta  S^{n^{\prime}n}(-x)\gamma^\alpha C S^{m^{\prime}mT}(-x)C \right\} \\
-\partial_\mu {\rm Tr}\! \left\{\gamma_\alpha S^{kk^{\prime}}(x)\gamma_\beta CS^{jj^{\prime}T}(x)C \right\}   \partial_\nu {\rm Tr}\! \left\{\gamma^\beta  S^{n^{\prime}n}(-x)\gamma^\alpha C S^{m^{\prime}mT}(-x)C \right\} \\
-\partial_\nu {\rm Tr} \!\left\{\gamma_\alpha S^{kk^{\prime}}(x)\gamma_\beta CS^{jj^{\prime}T}(x)C \right\}  \partial_\mu  {\rm Tr}\!\left\{\gamma^\beta  S^{n^{\prime}n}(-x)\gamma^\alpha C S^{m^{\prime}mT}(-x)C \right\} \\
+\partial_\mu \partial_\nu {\rm Tr}\! \left\{\gamma_\alpha S^{kk^{\prime}}(x)\gamma_\beta CS^{jj^{\prime}T}(x)C \right\}  {\rm Tr}\!\left\{\gamma^\beta  S^{n^{\prime}n}(-x)\gamma^\alpha C S^{m^{\prime}mT}(-x)C \right\}
\end{array}\right] \, , \nonumber\\
\end{eqnarray}
after performing the Wick's contractions, where the $S_{ij}(x)$ is the full $s$-quark propagator \cite{QCDSR-Reinders85,Pascual-1984,WangHuang3900},
\begin{eqnarray}\label{squark}
S_{ij}(x)&=& \frac{i\delta_{ij}\!\not\!{x}}{ 2\pi^2x^4}
-\frac{\delta_{ij}m_s}{4\pi^2x^2}-\frac{\delta_{ij}\langle
\bar{s}s\rangle}{12} +\frac{i\delta_{ij}\!\not\!{x}m_s
\langle\bar{s}s\rangle}{48}-\frac{\delta_{ij}x^2\langle \bar{s}g_s\sigma Gs\rangle}{192}+\frac{i\delta_{ij}x^2\!\not\!{x} m_s\langle \bar{s}g_s\sigma
 Gs\rangle }{1152}\nonumber\\
&& -\frac{ig_s G^{a}_{\alpha\beta}t^a_{ij}(\!\not\!{x}
\sigma^{\alpha\beta}+\sigma^{\alpha\beta} \!\not\!{x})}{32\pi^2x^2}  -\frac{1}{8}\langle\bar{s}_j\sigma^{\mu\nu}s_i \rangle \sigma_{\mu\nu}+\cdots \, ,
\end{eqnarray}
and  $t^n=\frac{\lambda^n}{2}$, the $\lambda^n$ is the Gell-Mann matrix,  we retain the term $\langle\bar{s}_j\sigma_{\mu\nu}s_i \rangle$ originates from the Fierz re-arrangement of the $\langle s_i \bar{s}_j\rangle$ to  absorb the gluons  emitted from other  quark lines   to extract the mixed condensate $\langle\bar{s}g_s\sigma G s\rangle$ \cite{WangHuang3900}.
 Then it is straightforward to  carry out the integrals $d^4x$ in the coordinate  space by setting $d^4x \to d^Dx$ and using the basic integral,
 \begin{eqnarray}
 \int d^D x \frac{e^{ip\cdot x}}{x^{2n}}&=&i(-1)^{n+1}\frac{2^{D-2n}\pi^{\frac{D}{2}}\Gamma\left( \frac{D}{2}-n\right)}{\Gamma(n)(-p^2)^{\frac{D}{2}-n}} \, ,
 \end{eqnarray}
    to obtain the correlation function $\Pi_{\mu\nu}(p)$, therefore we obtain the spectral representation at the quark-gluon level through dispersion relation,
 \begin{eqnarray}
 \Pi_Y(p^2)&=&\frac{1}{\pi}\int_0^\infty ds \frac{{\rm Im }\Pi_Y(s)}{s-p^2} \, ,
 \end{eqnarray}
the QCD spectral density $\rho_{QCD}(s)=\frac{{\rm Im }\Pi_Y(s)}{\pi} $, where we have used the formula,
\begin{eqnarray}
\frac{1}{\pi}{\rm Im}\frac{\Gamma(\epsilon-\alpha)}{(-p^2)^{\epsilon-\alpha}}&=&\frac{p^{2\alpha}}{n!}\, ,
\end{eqnarray}
$D=4+2\epsilon$, $\alpha=n-2$.

Then we take the simple replacement $\delta_{ij}\partial_\mu \to \delta_{ij}\partial_\mu-ig_sG^a_\mu \frac{\lambda^a_{ij}}{2}$ in the vertexes in Eq.\eqref{example} to take account of  additional terms originating  from the covariant derivatives, thus we obtain the corresponding correlation function for the gauge invariant current,   the calculation is straightforward via the same procedure.
The mass of the $s$-quark is taken as a small quantity and treated perturbatively in  Eq.\eqref{squark}, which  requires that the lower bound of the integral $ds$ should be zero.

When we acquire the analytical expressions  of the QCD spectral densities $\rho_{QCD}(s)$, we take the quark-hadron duality below the continuum thresholds $s_0$ by setting,
\begin{eqnarray}
\frac{1}{\pi} \int_0^{s_0} \frac{{\rm Im}\Pi_{Y}(s)}{s-p^2} &=& \int_0^{s_0} ds \frac{\rho_{QCD}(s)}{s-p^2}\, ,
\end{eqnarray}
and implement the Borel transform in regard to the variable $P^2=-p^2$  to get the QCD sum rules,
\begin{eqnarray}\label{SR-QCD}
\lambda^2_{Y}\, \exp\left(-\frac{M^2_{Y}}{T^2}\right)= \int_{0}^{s_0} ds\, \rho_{QCD}(s) \, \exp\left(-\frac{s}{T^2}\right) \, ,
\end{eqnarray}
the explicit expressions of the spectral densities $\rho_{QCD}(s)$ for the currents with both partial derivatives and covariant derivatives are all given in the Appendix.
The vacuum condensates $\langle \bar{q}q\rangle^3$ make no contribution due to
 the special structures of the currents, where $q=u$, $d$ or $s$.

We calculate the vacuum condensates with dimensions up to 11 in the operator product expansion, and  consider the impact of the vacuum condensates which are vacuum expectations of the quark-gluon operators of the orders $\mathcal{O}( \alpha_s^{k})$ with $k\leq 1$, just like in our previous works \cite{WangHuang3900,Wang-tetraquark-QCDSR-2,WZG-CPC-Zcs}.
The vacuum condensates $\langle g_s^3f^{abc}G_aG_bG_c\rangle$,  $\langle\frac{\alpha_s}{\pi}GG\rangle^2$ and $\langle\overline{q}g_s\sigma Gq\rangle\langle\frac{\alpha_s}{\pi}GG\rangle$ are the vacuum expectations of the quark-gluon operators of the orders $\mathcal{O}(\alpha_s^{\frac{3}{2}})$, $\mathcal{O}(\alpha_s^{2})$ and $\mathcal{O}(\alpha_s^{\frac{3}{2}})$, respectively, and are neglected, where $q=u$, $d$ or $s$. Direct calculations indicate that those contributions are tiny in the QCD sum rules for the multiquark states \cite{WXW-EPJA}. The vacuum condensates  $\langle\overline{q}q\rangle^4$ are companied with the strong coupling constant $g_s^2$, also make tiny contributions and are neglected for simplicity.

We differentiate  Eq.\eqref{SR-QCD} with respect to  $\tau=\frac{1}{T^2}$, then eliminate the pole residues  $\lambda_{Y}$, and obtain the QCD sum rules for the masses of the fully-light vector tetraquark states,
\begin{eqnarray}
 M^2_{Y}&=& -\frac{\int_{0}^{s_0} ds\frac{d}{d \tau}\rho_{QCD}(s)\exp\left(-\tau s \right)}{\int_{0}^{s_0} ds \rho_{QCD}(s)\exp\left(-\tau s\right)}\, .
\end{eqnarray}

\section{Numerical results and discussions}

We take the standard values of the vacuum condensates $\langle
\bar{q}q \rangle=-(0.24\pm 0.01\, \rm{GeV})^3$,   $\langle
\bar{q}g_s\sigma G q \rangle=m_0^2\langle \bar{q}q \rangle$,
$m_0^2=(0.8 \pm 0.1)\,\rm{GeV}^2$, $\langle\bar{s}s \rangle=(0.8\pm0.1)\langle\bar{q}q \rangle$, $\langle\bar{s}g_s\sigma G s \rangle=m_0^2\langle \bar{s}s \rangle$,  $\langle \frac{\alpha_s GG}{\pi}\rangle=(0.012\pm0.004)\,\rm{GeV}^4 $  at the energy scale  $\mu=1\, \rm{GeV}$ \cite{QCDSR-SVZ79,QCDSR-Reinders85,QCDSR-Colangelo-Review}, and choose the $\overline{MS}$ mass $m_s(\mu=2\,\rm{GeV})=0.095\pm 0.005\,\rm{GeV}$ from the Particle Data Group \cite{PDG-2020}. We choose the square, cube and Nth powers of the $s$ quark mass $m_s^N=0$ $(N=2,3,4...)$ and neglect the small $u$ and $d$ quark masses.
 We usually choose the energy scale $\mu = 1\,\rm{GeV}$ for the QCD spectral densities of the fully-light mesons and baryons \cite{ssss-ZGWANG-2019,WZG-EPJC-Scalar,WZG-Vector-CPL}, and evolve the $s$-quark mass to the energy scale $\mu = 1\,\rm{GeV}$ accordingly.

 According to the experiential mass  gaps between the ground states and first radial excited states of the mesons,  we restrict the central values of the fully-light vector tetraquark masses $M_c$ and continuum threshold parameters ${\sqrt{s_0}}_c$ in the range about,
 \begin{eqnarray}
0.50 \,{\rm GeV} \leq {\sqrt{s_0}}_c - M_c \leq 0.60 \,{\rm GeV}\, ,
\end{eqnarray}
  in order to ensure uniformity and reliability of the results. For the conventional pseudoscalar and vector  mesons with the valence quarks $s$, $c$ and $b$, the mass gaps between the ground states and first radial excited states are about $0.55\sim 0.75\,\rm{GeV}$ from the Particle Data Group \cite{PDG-2020}. We borrow some idea from the experimental data and add an uncertainty $\delta {\sqrt{s_0}}=\pm 0.10\,\rm{GeV}$, then the continuum threshold parameters are about ${\sqrt{s_0}}={\sqrt{s_0}}_c\pm 0.10\,{\rm{GeV}}=M_c+0.40\sim0.70\,\rm{GeV}$, such as a strict constraint is reasonable.

We control the pole contributions (PC) range from $35\%$ to $75\%$ to ensure pole dominance at the  phenomenological  side whenever possible, and the PC are defined by
\begin{eqnarray}
\text{PC}&=&\frac{\int_{0}^{s_{0}}ds\,\rho_{QCD}\left(s\right)\exp\left(-\frac{s}{T^{2}}\right)} {\int_{0}^{\infty}ds\,\rho_{QCD}\left(s\right)\exp\left(-\frac{s}{T^{2}}\right)}\, .
\end{eqnarray}
The contributions of the vacuum condensates $D(n)$ in the operator product expansion are defined by
\begin{eqnarray}\label{Dn-define}
D(n)&=&\frac{\int_{0}^{s_{0}}ds\,\rho_{n}(s)\exp\left(-\frac{s}{T^{2}}\right)}
{\int_{0}^{s_{0}}ds\,\rho_{QCD}\left(s\right)\exp\left(-\frac{s}{T^{2}}\right)}\ ,
\end{eqnarray}
where the subscript $n$ in the QCD spectral densities $\rho_ n(s)$ denotes the dimension of the vacuum condensates,
\begin{eqnarray}
\rho_{3}(s)&\propto& \langle\bar{q}q\rangle\, ,\nonumber\\
\rho_{4}(s)&\propto& \langle\frac{\alpha_{s}GG}{\pi}\rangle\, ,\nonumber\\
\rho_{5}(s)&\propto& \langle\bar{q}g_s\sigma Gq\rangle\, ,\nonumber\\
\rho_{6}(s)&\propto& \langle\bar{q}q\rangle^2\, ,\nonumber\\
\rho_{7}(s)&\propto& \langle\bar{q}q\rangle\langle\frac{\alpha_{s}GG}{\pi}\rangle\, ,\nonumber\\
\rho_{8}(s)&\propto& \langle\bar{q}q\rangle\langle\bar{q}g_s\sigma Gq\rangle\, ,\nonumber\\
\rho_{9}(s)&\propto& \langle\bar{q}q\rangle^3 \, ,\nonumber\\
\rho_{10}(s)&\propto&  \langle\bar{q}g_s\sigma Gq\rangle^2\, , \langle\bar{q}q\rangle^2\langle\frac{\alpha_{s}GG}{\pi}\rangle\, ,\nonumber\\
\rho_{11}(s)&\propto&  \langle\bar{q}q\rangle^2\langle\bar{q}g_s\sigma Gq\rangle\, ,
\end{eqnarray}
where $q=u$, $d$ or $s$. We require the criterion $D(11)\sim 0\%$ to judge the convergent behaviors.
Then we get the suitable Borel parameters and continuum threshold parameters via trial and error.
In Table \ref{mass-residue-1}, we show the numerical results for the fully-light vector tetraquark states in the case of the partial derivatives.

From the analytical expressions of the QCD spectral densities shown in the Appendix, we can see explicitly that the currents with the partial derivatives and covariant derivatives  make no difference for the  tetraquark states  $Y^{1/3}_{ss\bar{s}\bar{s}}$, $Y^{1/3/5}_{qs\bar{q}\bar{s}}$ and $Y^{1/3}_{qq\bar{q}\bar{q}}$. On the other hand, for the tetraquark states $Y^2_{ss\bar{s}\bar{s}}$, $Y^2_{qs\bar{q}\bar{s}}$, $Y^2_{qq\bar{q}\bar{q}}$, $Y^{4}_{ss\bar{s}\bar{s}}$, $Y^{4/6}_{qs\bar{q}\bar{s}}$ and $Y^{4}_{qq\bar{q}\bar{q}}$, the currents with   covariant derivatives lead to slightly/tinily  different masses or pole residues comparing with the corresponding ones with partial derivatives. All the  numerical results  with the covariant derivatives would be  presented  in Table \ref{mass-residue-2}.

\begin{figure}
 \centering
  \includegraphics[totalheight=6cm,width=7cm]{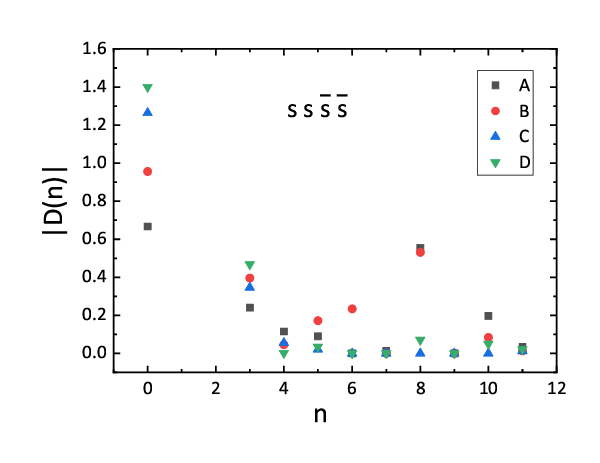}
  \includegraphics[totalheight=6cm,width=7cm]{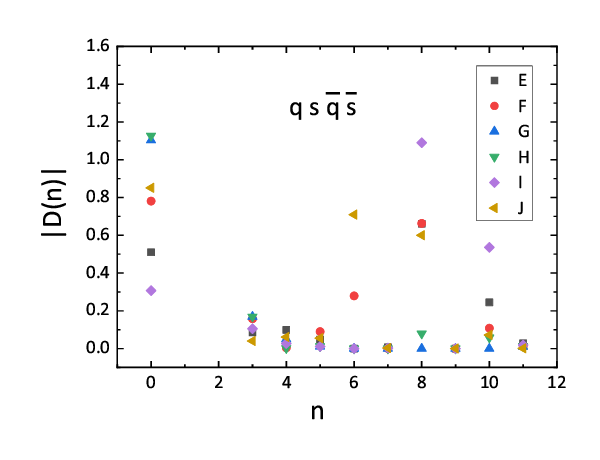}
  \includegraphics[totalheight=6cm,width=7cm]{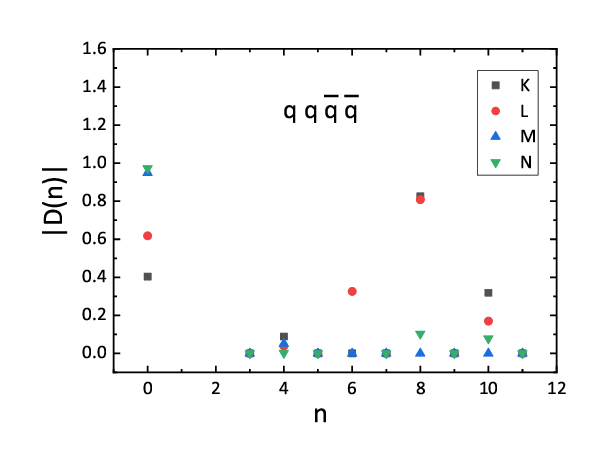}
     \caption{The absolute contributions of the vacuum condensates of dimension $n$ for the central values of the  input  parameters for the $ss\bar{s}\bar{s}$ and $qs\bar{q}\bar{s}$ P-wave (with partial derivatives) fully-light vector tetraquark states, where the $A$, $B$, $C$ and $D$ denote the  $Y^1_{ss\bar{s}\bar{s}}$, $Y^2_{ss\bar{s}\bar{s}}$, $Y^3_{ss\bar{s}\bar{s}}$ and $Y^4_{ss\bar{s}\bar{s}}$, the $E$, $F$, $G$, $H$, $I$ and $J$ denote the  $Y^1_{qs\bar{q}\bar{s}}$, $Y^2_{qs\bar{q}\bar{s}}$, $Y^3_{qs\bar{q}\bar{s}}$, $Y^4_{qs\bar{q}\bar{s}}$, $Y^5_{qs\bar{q}\bar{s}}$ and $Y^6_{qs\bar{q}\bar{s}}$, the $K$, $L$, $M$ and $N$ denote the  $Y^1_{qq\bar{q}\bar{q}}$, $Y^2_{qq\bar{q}\bar{q}}$, $Y^3_{qq\bar{q}\bar{q}}$ and $Y^4_{qq\bar{q}\bar{q}}$, respectively.}\label{Dn-ss-qs}
\end{figure}

In calculations, taking the currents with the partial derivatives as an example, we observe that the perturbative terms $D(0)$, $D(3)$, $D(6)$, $D(8)$ and $D(10)$ provide the most significant contributions, see Fig.\ref{Dn-ss-qs}. In general, although the contributions vary (remarkably)   with the vacuum condensates of increasing dimensions, the contributions $D(0)$ or $D(8)$ serve as milestones, the higher dimensional vacuum condensates play a less (or decreasing) important role, and $D(11)\sim 0\%$. More explicitly, for the curves $C$, $D$, $G$, $H$, $M$ and $N$, the dominant contributions come from the $D(0)$; for the curves $A$, $B$, $F$ and $J$, the largest contributions come from the $D(0)$, the contributions from the $D(8)$ are also large, then the higher vacuum condensate contributions decrease quickly to zero; for the curves $E$, $I$, $K$ and $L$, the largest contributions come from the $D(8)$,  then the higher vacuum condensate contributions decrease quickly to zero. All in all, for the curves $A$, $B$, $C$, $D$, $F$, $G$, $H$, $J$, $M$ and $N$, the operator product expansion converges very good; for the curves $E$, $I$, $K$ and $L$, the operator product expansion converges.

\begin{table}
\begin{center}
\begin{tabular}{|c|c|c|c|c|c|c|c|c|}\hline\hline
$Y$    & $T^2 (\rm{GeV}^2)$   &$\sqrt{s_0}(\rm GeV) $    &pole
&$M (\rm{GeV})$   & $\lambda (\rm{GeV}^6) $
\\ \hline

$Y^1_{ss\bar{s}\bar{s}}$     & $1.30-1.60$            &$2.75\pm0.10$   &$(37-70)\%$                       & $2.16\pm0.14$              & $(2.67\pm0.50)\times 10^{-2}$                             \\

$Y^2_{ss\bar{s}\bar{s}}$     & $1.45-1.75$           &$2.95\pm0.10$    &$(35-66)\%$
& $2.35\pm0.17$              & ($3.67\pm0.89)\times 10^{-2}$                             \\

$Y^3_{ss\bar{s}\bar{s}}$     & $1.75-2.15$           &$3.50\pm0.10$    &$(42-71)\%$
& $2.98\pm0.10$              & $(2.86\pm0.43)\times 10^{-1}$                              \\

$Y^4_{ss\bar{s}\bar{s}}$     &  $1.80-2.20$          &$3.65\pm0.10$    &$(46-74)\%$
&$3.13\pm0.11$               & $(1.58\pm0.22)\times 10^{-1}$                                 \\
\hline

$Y^1_{qs\bar{q}\bar{s}}$     & $1.25-1.55$           &$2.65\pm0.10$    &$(38-71)\%$
& $2.06\pm0.13$              & $(1.19\pm0.23)\times 10^{-2}$                               \\

$Y^2_{qs\bar{q}\bar{s}}$     & $1.40-1.70$           &$2.85\pm0.10$    &$(35-67)\%$
&$2.24\pm0.17$               & $(1.90\pm0.46)\times 10^{-2}$                                 \\

$Y^3_{qs\bar{q}\bar{s}}$     & $1.70-2.10$            &$3.40\pm0.10$   &$(41-71)\%$
& $2.87\pm0.10$              & $(1.24\pm0.19)\times 10^{-1}$                             \\

$Y^4_{qs\bar{q}\bar{s}}$     & $1.75-2.15$            &$3.55\pm0.10$   &$(46-75)\%$
&$ 3.00\pm0.11$              & $(6.77\pm1.01)\times 10^{-2}$
\\

$Y^5_{qs\bar{q}\bar{s}}$     & $1.30-1.70$            &$2.60\pm0.10$    &$(37-73)\%$
& $2.05\pm0.10$              & $(5.79\pm0.64)\times 10^{-3}$                               \\

$Y^6_{qs\bar{q}\bar{s}}$     & $2.50-3.40$            &$4.05\pm0.10$    &$(44-74)\%$
&$3.52\pm0.11$               & $(3.89\pm0.85)\times 10^{-2}$
\\
\hline

$Y^1_{qq\bar{q}\bar{q}}$     & $1.20-1.50$           &$2.55\pm0.10$     &$(39-72)\%$             & $1.98\pm0.11$              & $(2.15\pm0.33)\times 10^{-2}$                              \\

$Y^2_{qq\bar{q}\bar{q}}$     & $1.35-1.65$           &$2.75\pm0.10$     &$(36-68)\%$
& $2.13\pm0.16$              & $(2.84\pm0.58)\times 10^{-2}$                                       \\

$Y^3_{qq\bar{q}\bar{q}}$     & $1.65-2.05$           &$3.30\pm0.10$     &$(40-71)\%$      & $2.77\pm0.11$              & $(2.18\pm0.33)\times 10^{-1}$                             \\

$Y^4_{qq\bar{q}\bar{q}}$     & $1.70-2.10$           &$3.45\pm0.10$     &$(46-75)\%$
& $2.89\pm0.10$              & $(1.21\pm0.17)\times 10^{-1}$                                \\

\hline\hline
\end{tabular}
\end{center}
\caption{ The Borel parameters, continuum threshold parameters, pole contributions, masses and pole residues of the P-wave (with partial derivatives) fully-light vector tetraquark states. }\label{mass-residue-1}
\end{table}

\begin{table}
\begin{center}
\begin{tabular}{|c|c|c|c|c|c|c|c|c|}\hline\hline
$Y$    & $T^2 (\rm{GeV}^2)$   &$\sqrt{s_0}(\rm GeV) $    &pole
&$M (\rm{GeV})$   & $\lambda (\rm{GeV}^6) $
\\ \hline

$Y^1_{ss\bar{s}\bar{s}}$     & $1.30-1.60$            &$2.75\pm0.10$   &$(37-70)\%$                       & $2.16\pm0.14$              & $(2.67\pm0.50)\times 10^{-2}$
\\

$Y^2_{ss\bar{s}\bar{s}}$     & $1.45-1.75$           &$2.95\pm0.10$    &$(35-65)\%$
& $2.35\pm0.17$              & ($3.64\pm0.83)\times 10^{-2}$                             \\

$Y^3_{ss\bar{s}\bar{s}}$     & $1.75-2.15$           &$3.50\pm0.10$    &$(42-71)\%$
& $2.98\pm0.10$              & $(2.86\pm0.43)\times 10^{-1}$
\\

$Y^4_{ss\bar{s}\bar{s}}$     &  $1.80-2.20$          &$3.65\pm0.10$    &$(45-73)\%$
&$3.16\pm0.10$               & $(1.60\pm0.22)\times 10^{-1}$
\\    \hline

$Y^1_{qs\bar{q}\bar{s}}$     & $1.25-1.55$           &$2.65\pm0.10$    &$(38-71)\%$
& $2.06\pm0.13$              & $(1.19\pm0.23)\times 10^{-2}$                               \\

$Y^2_{qs\bar{q}\bar{s}}$     & $1.40-1.70$           &$2.85\pm0.10$    &$(35-67)\%$
&$2.24\pm0.17$               & $(1.88\pm0.38)\times 10^{-2}$                                 \\

$Y^3_{qs\bar{q}\bar{s}}$     & $1.70-2.10$            &$3.40\pm0.10$   &$(41-71)\%$
& $2.87\pm0.10$              & $(1.24\pm0.19)\times 10^{-1}$                             \\

$Y^4_{qs\bar{q}\bar{s}}$     & $1.75-2.15$            &$3.55\pm0.10$   &$(45-74)\%$
&$ 3.02\pm0.11$              & $(6.89\pm0.95)\times 10^{-2}$
\\

$Y^5_{qs\bar{q}\bar{s}}$     & $1.30-1.70$            &$2.60\pm0.10$    &$(37-73)\%$
& $2.05\pm0.10$              & $(5.79\pm0.64)\times 10^{-3}$                               \\

$Y^6_{qs\bar{q}\bar{s}}$     & $2.50-3.40$            &$4.05\pm0.10$    &$(44-74)\%$
&$3.54\pm0.09$               & $(3.93\pm0.74)\times 10^{-2}$
\\
\hline

$Y^1_{qq\bar{q}\bar{q}}$     & $1.20-1.50$           &$2.55\pm0.10$     &$(39-72)\%$             & $1.98\pm0.11$              & $(2.15\pm0.33)\times 10^{-2}$                              \\

$Y^2_{qq\bar{q}\bar{q}}$     & $1.35-1.65$           &$2.75\pm0.10$     &$(35-67)\%$
& $2.13\pm0.16$              & $(2.78\pm0.51)\times 10^{-2}$                                       \\

$Y^3_{qq\bar{q}\bar{q}}$     & $1.65-2.05$           &$3.30\pm0.10$     &$(40-71)\%$      & $2.77\pm0.11$              & $(2.18\pm0.33)\times 10^{-1}$                             \\

$Y^4_{qq\bar{q}\bar{q}}$     & $1.70-2.10$           &$3.45\pm0.10$     &$(45-75)\%$
& $2.90\pm0.10$              & $(1.21\pm0.16)\times 10^{-1}$                              \\

\hline\hline
\end{tabular}
\end{center}
\caption{ The Borel parameters, continuum threshold parameters, pole contributions, masses and pole residues of the P-wave  (with covariant derivatives) fully-light vector tetraquark states. }\label{mass-residue-2}
\end{table}

It is straightforward to obtain the masses and pole residues of the vector tetraquark states, which are shown explicitly in Tables \ref{mass-residue-1}-\ref{mass-residue-2}, where we calculate the uncertainties $\delta$  with the
formula,
\begin{eqnarray}
\delta=\sqrt{\sum_i\left(\frac{\partial f}{\partial
x_i}\right)^2\mid_{x_i=\bar{x}_i} (x_i-\bar{x}_i)^2}\,  ,
\end{eqnarray}
 where the $f$ denote  the tetraquark masses $M_Y$ and pole residues $\lambda_Y$,  the $x_i$
denote all the input parameters $m_s$, $\langle \bar{q}q
\rangle$, $\langle \bar{s}s \rangle$, $\cdots$. As the partial
 derivatives   $\frac{\partial f}{\partial x_i}$ are difficult to carry
out analytically, we take the  approximation $\left(\frac{\partial
f}{\partial x_i}\right)^2 (x_i-\bar{x}_i)^2\approx
\left[f(\bar{x}_i\pm \delta x_i)-f(\bar{x}_i)\right]^2$.

In calculations, we observe that if we choose the continuum threshold parameter $\sqrt{s_0}=2.75\pm 0.10\,\rm{GeV}$ and Borel parameter $T^2=(1.3-1.6)\,\rm{GeV}^2$, then the pole contribution is about $(37-70)\%$ for the vector tetraquark state $Y^1_{ss\bar{s}\bar{s}}$, the predicted mass $M = 2.16 \pm 0.14\,\rm{GeV}$,
see Tables \ref{mass-residue-1}-\ref{mass-residue-2},  matches the experimental value $M_Y  = 2.175 \pm 0.010\pm 0.015\,\rm{GeV}$ for the $Y(2175)/\phi(2170)$ \cite{BabarY2175-2006}, and supports assigning the $Y(2175)/\phi(2170)$ as a vector $ss\bar{s}\bar{s}$ tetraquark state.

In this paper, we choose the pole contributions about $(35-75)\%$ consistently, the largest pole contributions up to know, just like in our previous work \cite{ssss-ZGWANG-2019}, where the S-wave diquark-antidiquark type $ss\bar{s}\bar{s}$ and $qq\bar{q}\bar{q}$  states are explored with the QCD sum rules. In the QCD sum rules, we extract the masses in the Borel windows, which depend on the pole contributions and convergent behaviors of the operator product expansion, the predicted masses change  with variations of the Borel windows. There are two choices  in defining the contributions of the vacuum condensates $D(n)$, the definition in Eq.\eqref{Dn-define} is chosen in the present work and Refs.\cite{Y2175CHX2008,Y2175CHX2018,ssss-ZGWANG-2019}, while in Ref.\cite{Y2175CHX2022}, the $D(n)$ is defined by setting $s_0 \to \infty$. It is not odd that the existing  predictions are compatible or slightly different.  In Refs.\cite{Y2175CHX2008,Y2175CHX2018}, Chen et al obtain the masses $2.41 \pm 0.25 \,\rm{GeV}$ and $2.34 \pm 0.17\,\rm{GeV}$ for the two lowest vector tetraquark states $ss\bar{s}\bar{s}$, the central values are larger than that of the present calculation $ 2.16 \pm 0.14\,\rm{GeV}$. Furthermore, we should bear in mind, the interpolating currents have the same quantum numbers $J^{PC}$ but different structures, which correspond to several tetraquark states or a tetraquark state
with several Fock components, it is not odd that the predictions are also different.

 Another noteworthy point is that the masses of the $Y^2_{qs\bar{q}\bar{s}}$ and $Y^2_{ss\bar{s}\bar{s}}$, see  Tables \ref{mass-residue-1}-\ref{mass-residue-2},  are  $M=2.24 \pm 0.17\,\rm{GeV}$ and $ 2.35 \pm 0.17\,\rm{GeV}$ respectively, which happen to fit the experimentally observed states $X(2240)$ and $X(2400)$ with the masses in the range $(2.20-2.40)\,\rm{GeV}$ \cite{BESIIIY2175(2239)-2018,BESIII2232-2019,BABAR2232-2020,BESIII2356-2022}. We think it is possible that the $X(2240)$ and $X(2400)$ are the $qs\bar{q}\bar{s}$ and $ss\bar{s}\bar{s}$   states respectively with the structure $C\gamma_\mu\otimes \stackrel{\leftrightarrow}{\partial}_\alpha \otimes\gamma^\alpha C + C\gamma^\alpha \otimes\stackrel{\leftrightarrow}{\partial}_\alpha \otimes\gamma_\mu$ or $C\gamma_\mu\otimes \stackrel{\leftrightarrow}{D}_\alpha \otimes\gamma^\alpha C + C\gamma^\alpha \otimes\stackrel{\leftrightarrow}{D}_\alpha \otimes\gamma_\mu$. The assignments are consistent with the observation
of the $X(2240)$ in the $K^+K^-$ invariant mass spectrum \cite{BESIIIY2175(2239)-2018}, as the decay $Y^2_{qs\bar{q}\bar{s}} \to K^+K^-$ can take place through the Okubo-Zweig-Iizuka super-allowed fall-apart mechanism. While the decay $Y^2_{ss\bar{s}\bar{s}} \to \Lambda \bar{\Lambda}$ could take place through annihilating an $s\bar{s}$ pair and creating a $u\bar{u}$ pair and a $d\bar{d}$ pair, which is not  Okubo-Zweig-Iizuka favored, we can search for the two-body strong decays $Y^2_{ss\bar{s}\bar{s}} \to \phi f_0(980)$, $\phi \eta^\prime$ to diagnose the nature of the $X(2400)$.

In our previous work \cite{ssss-ZGWANG-2019}, the predicted mass of the vector tetraquark state without an explicit P-wave between the diquark and antidiquark pair, $M_{X}=3.08\pm0.11\,\rm{GeV}$, is much larger than the experimental value of the mass of the $Y(2175)/\phi(2170)$, and opposes assigning  the $Y(2175)/\phi(2170)$ as a hidden-strange partner of the $Y$ states without an explicit P-wave.
Most of the vector tetraquark states with an explicit P-wave have lower masses than the corresponding vector tetraquark states with an  implicit P-wave we studied earlier \cite{ssss-ZGWANG-2019}, see Tables  \ref{mass-residue-1}-\ref{mass-residue-2}. If we release the pole contributions $(37-70)\%$ to smaller values, we can obtain even smaller masses than those presented  in Tables  \ref{mass-residue-1}-\ref{mass-residue-2}, for example, the values obtained  in Refs.\cite{ssss-ZGWANG-2006,WZG-Vector-CPL}.

\begin{figure}
 \centering
  \includegraphics[totalheight=6cm,width=7cm]{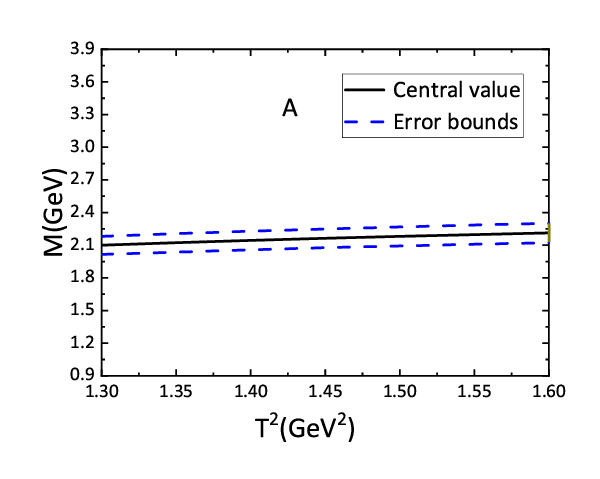}
  \includegraphics[totalheight=6cm,width=7cm]{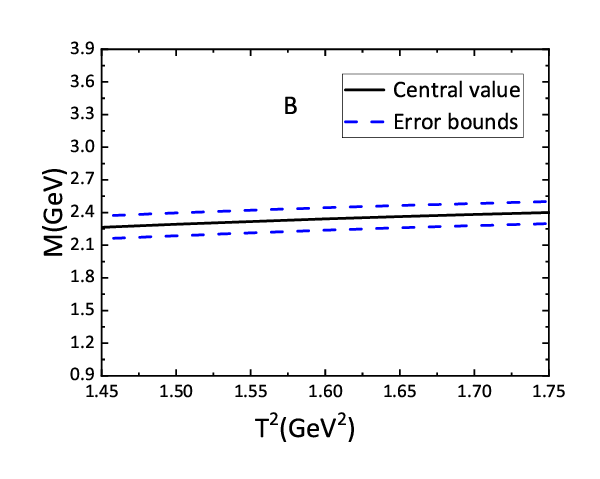}
  \includegraphics[totalheight=6cm,width=7cm]{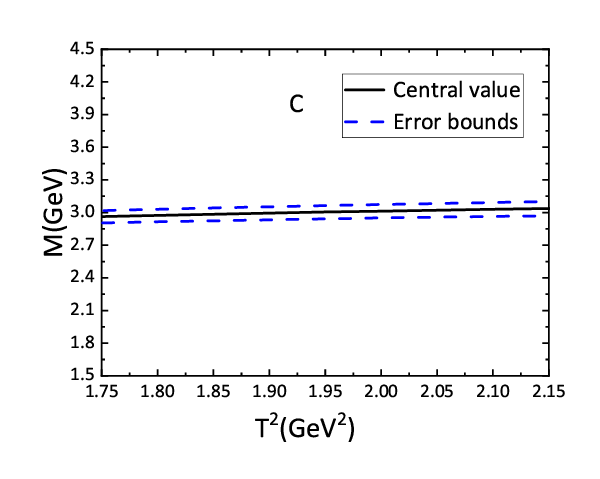}
  \includegraphics[totalheight=6cm,width=7cm]{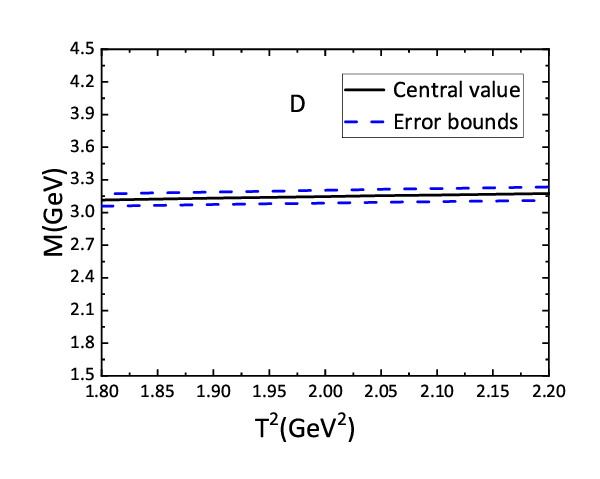}
     \caption{The masses  with variations of the  Borel  parameters  $T^2$, where the $A$, $B$, $C$, and $D$ denote the  $Y^1_{ss\bar{s}\bar{s}}$, $Y^2_{ss\bar{s}\bar{s}}$, $Y^3_{ss\bar{s}\bar{s}}$ and $Y^4_{ss\bar{s}\bar{s}}$ P-wave (with partial derivatives) fully-light vector tetraquark states, respectively.}\label{mass-ssss}
\end{figure}

\begin{figure}
 \centering
  \includegraphics[totalheight=6cm,width=7cm]{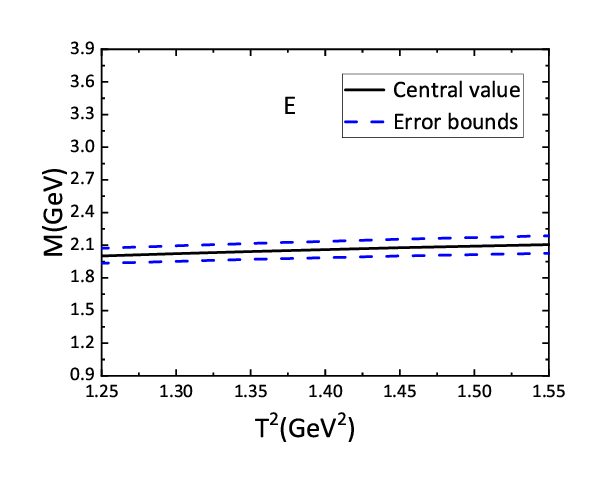}
  \includegraphics[totalheight=6cm,width=7cm]{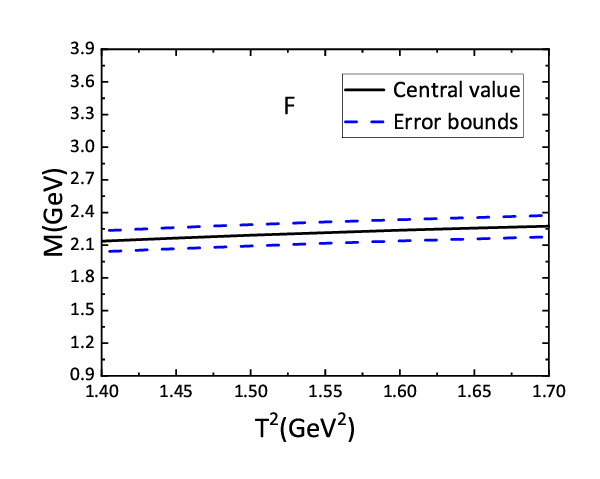}
  \includegraphics[totalheight=6cm,width=7cm]{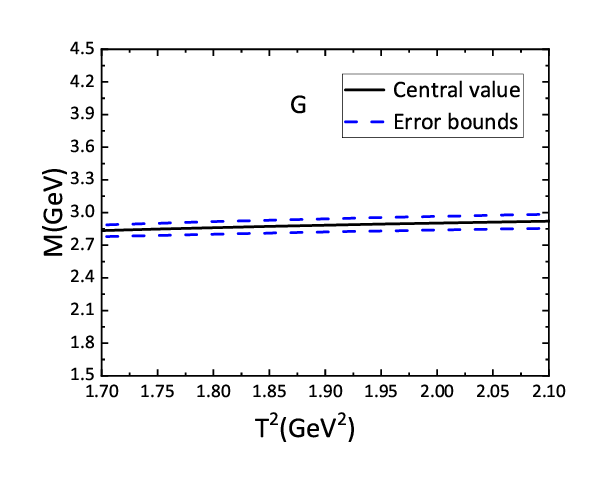}
  \includegraphics[totalheight=6cm,width=7cm]{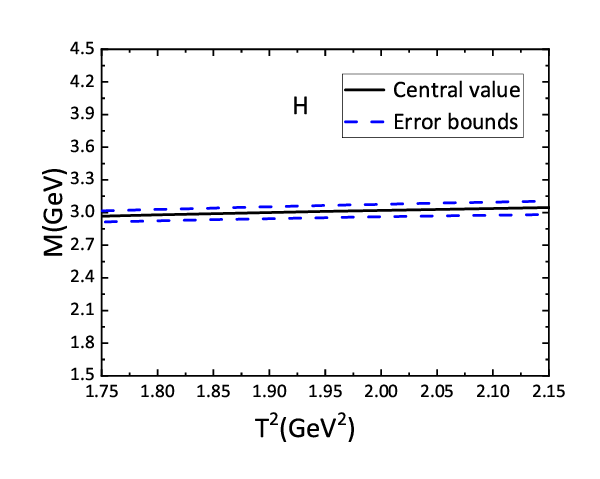}
  \includegraphics[totalheight=6cm,width=7cm]{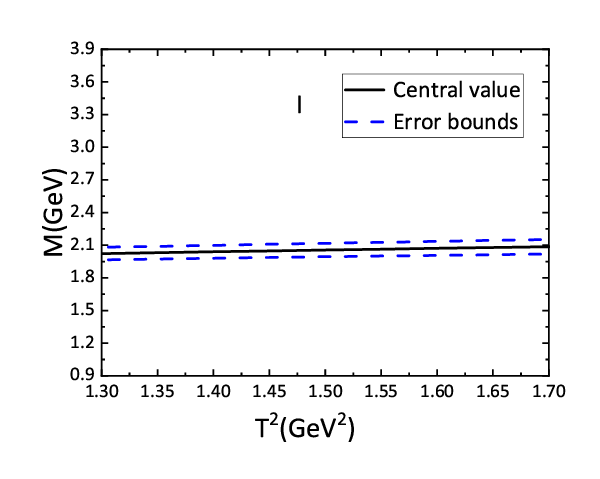}
  \includegraphics[totalheight=6cm,width=7cm]{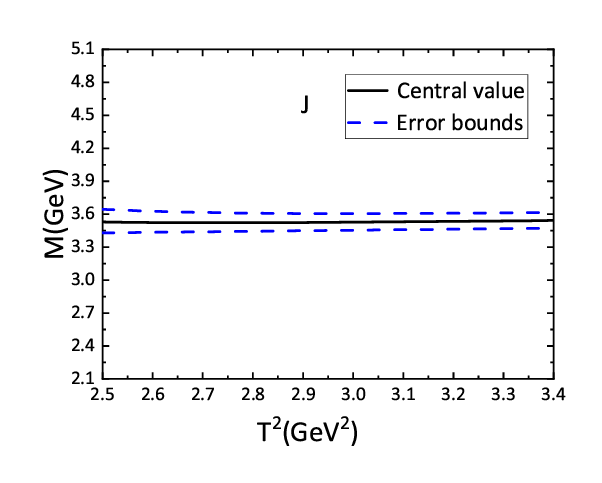}
     \caption{The masses  with variations of the  Borel  parameters  $T^2$, where the $E$, $F$, $G$, $H$, $I$ and $J$ denote the $Y^1_{qs\bar{q}\bar{s}}$, $Y^2_{qs\bar{q}\bar{s}}$, $Y^3_{qs\bar{q}\bar{s}}$, $Y^4_{ss\bar{s}\bar{s}}$,$Y^5_{qs\bar{q}\bar{s}}$ and $Y^6_{qs\bar{q}\bar{s}}$ P-wave (with partial derivatives) fully-light vector tetraquark states, respectively.}\label{mass-qqss}
\end{figure}

\begin{figure}
 \centering
  \includegraphics[totalheight=6cm,width=7cm]{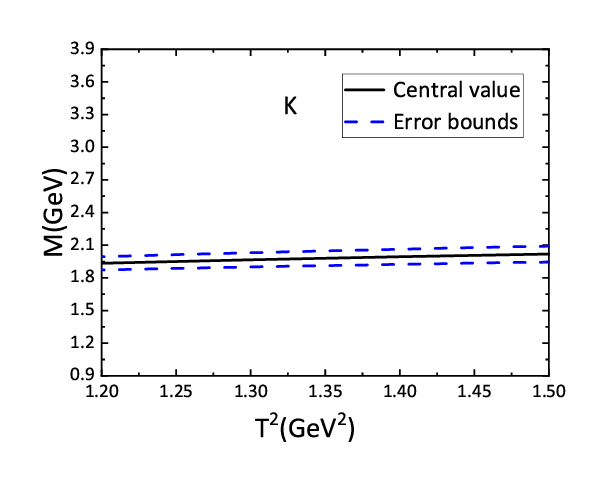}
  \includegraphics[totalheight=6cm,width=7cm]{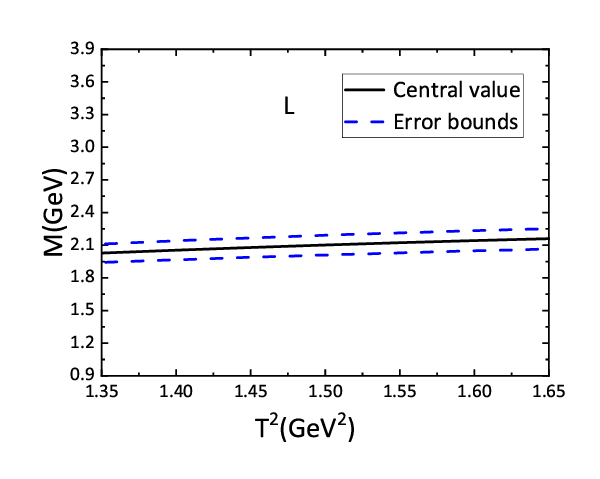}
  \includegraphics[totalheight=6cm,width=7cm]{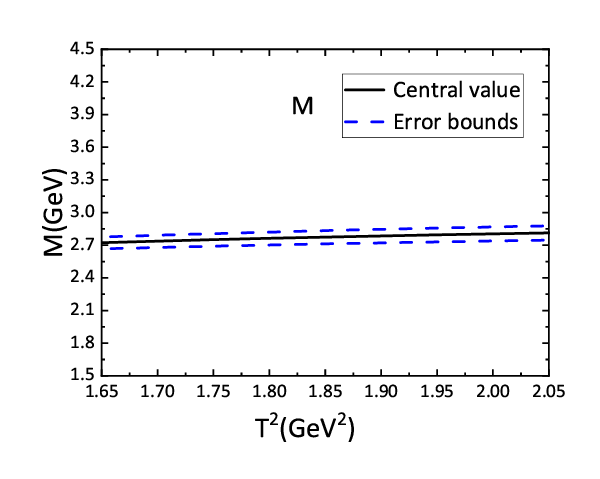}
  \includegraphics[totalheight=6cm,width=7cm]{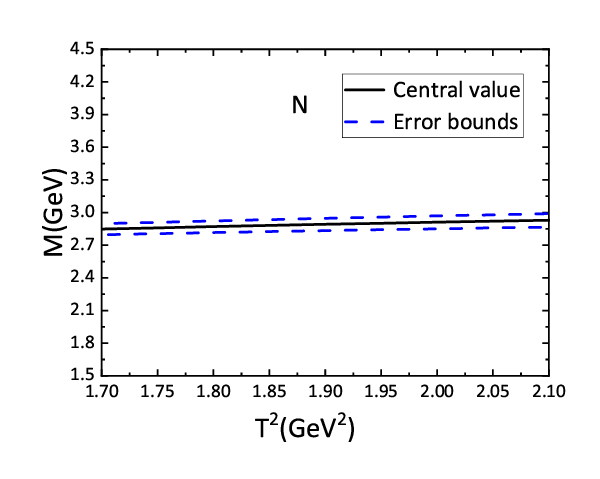}
     \caption{The masses  with variations of the  Borel  parameters  $T^2$, where the $K$, $L$, $M$ and $N$ denote the $Y^1_{qq\bar{q}\bar{q}}$, $Y^2_{qq\bar{q}\bar{q}}$, $Y^3_{qq\bar{q}\bar{q}}$ and $Y^4_{qq\bar{q}\bar{q}}$  P-wave (with partial derivatives) fully-light vector tetraquark states, respectively.}\label{mass-qqqq}
\end{figure}

The predicted masses are stable with variations of the Borel parameters, see Figs.\ref{mass-ssss}-\ref{mass-qqqq} for the currents with partial derivatives  as an example, the uncertainties come from the Borel parameters are rather small, the predictions are robust. Furthermore, from Tables \ref{mass-residue-1}-\ref{mass-residue-2}, we can see clearly  that the currents with the covariant derivatives and partial derivatives only lead to slight/tiny difference, we prefer the  covariant derivatives in constructing the interpolating currents if gauge invariance is emphasized.
On the other hand, if only final numerical results are concerned, we can choose either covariant derivatives or partial derivatives.

The currents $J_{\alpha}$ with the same quantum numbers could mix with each other under renormalization, we should introduce the mixing matrixes $U$ to obtain the diagonal currents, $J_{\alpha}^{\prime }=UJ_{\alpha}$, which couple potentially to (more) physical states, as the physical states  have several Fock components.  The matrixes $U$ can be determined by direct calculating anomalous dimensions of all the currents,  it is difficult to  obtain a diagonal current, which is an special superposition of several non-trial currents to match with all the considerable Fock states. For the vector tetraquark state $ss\bar{s}\bar{s}$, if we want to obtain a more physical state, we have to take account of the mixing effects of the currents $J^1_{\mu,ss\bar{s}\bar{s}}$, $J^2_{\mu,ss\bar{s}\bar{s}}$, $J^3_{\mu,ss\bar{s}\bar{s}}$ and $J^4_{\mu,ss\bar{s}\bar{s}}$ at least, and introduce three mixing angles, $\theta$, $\theta_{12}$ and $\theta_{34}$,
\begin{eqnarray}
J_{\mu,ss\bar{s}\bar{s}}&=&\cos\theta  \left( \cos\theta_{12} J^1_{\mu,ss\bar{s}\bar{s}}+\sin\theta_{12} J^2_{\mu,ss\bar{s}\bar{s}}\right)+\sin\theta \left(
\cos\theta_{34} J^3_{\mu,ss\bar{s}\bar{s}}+\sin\theta_{34} J^4_{\mu,ss\bar{s}\bar{s}}\right)\, ,
\end{eqnarray}
it is a difficult work. The currents $J^1_{\mu,ss\bar{s}\bar{s}}$, $J^2_{\mu,ss\bar{s}\bar{s}}$, $J^3_{\mu,ss\bar{s}\bar{s}}$ and $J^4_{\mu,ss\bar{s}\bar{s}}$ maybe couple potentially to four different  tetraquark states or to a tetraquark state with four different Fock components. Without exploring the mixing effects and strong decays in details combined with precise experimental data, we cannot assign the $Y(2175)/\phi(2170)$, $X(2240)$, $X(2400)$ in a
rigorous way. At the present time, we just assign those exotic states tentatively and roughly. According to the BESIII and BaBar data, the $Y(2175)/\phi(2170)$ and $X(2240)$ are two different states \cite{BESIIIY2175(2239)-2018,BABAR2232-2020}, their
valence quark constituents are favored to be $ss\bar{s}\bar{s}$ and $qs\bar{q}\bar{s}$, respectively, the mixing effects between the currents $J^1_{\mu,ss\bar{s}\bar{s}}$ and $J^2_{\mu,qs\bar{q}\bar{s}}$ are expected to be small,  the present assignments  make sense.
We hope those vector tetraquark states can be experimentally verified in the future, and confront the predictions to the experimental data in the future at the BESIII, LHCb, Belle II et al.

\section{Conclusion}
In this paper, we construct the fully-light vector/tensor four-quark currents and introduce an explicit P-wave between the diquark and antidiquark pair. We take account of the contributions of the vacuum condensates up to dimension 11 in the operator product expansion and obtain the QCD sum rules for the masses and pole resides of the vector tetraquark states. For one of the $ss\bar{s}\bar{s}$ structures, the predicted mass $M = 2.16\pm0.14\,\rm{GeV}$ is in great agreement with the experimental value of the $Y(2175)/\phi(2170)$ from the BaBar/BESIII/Belle collaborations, and supports assigning the $Y(2175)/\phi(2170)$  as the $C\gamma_\alpha\otimes\stackrel{\leftrightarrow}{\partial}_\mu\otimes\gamma_\alpha C$-type (or $C\gamma_\alpha\otimes\stackrel{\leftrightarrow}D_\mu\otimes\gamma^\alpha C$-type) fully-strange vector tetraquark state with an explicit P-wave. The  $C\gamma_\mu\otimes \stackrel{\leftrightarrow}{\partial}_\alpha \otimes\gamma^\alpha C + C\gamma^\alpha \otimes\stackrel{\leftrightarrow}{\partial}_\alpha \otimes\gamma_\mu$ (or $C\gamma_\mu\otimes \stackrel{\leftrightarrow}D_\alpha \otimes\gamma^\alpha C + C\gamma^\alpha \otimes\stackrel{\leftrightarrow}D_\alpha \otimes\gamma_\mu$) type  $qs\bar{q}\bar{s} $ and $ss\bar{s}\bar{s}$   vector tetraquark states have masses which are compatible with the experimental values of  the $X(2240)$ and $X(2400)$ respectively in the region $2.20$ to $2.40\,\rm{GeV}$ from the BaBar/BESIII  collaborations. All in all, we predict that the central values of the  masses of the vector $ss\bar{s}\bar{s}$ tetraquark states with an explicit P-wave lie at the region $2.16-3.13\,\rm{GeV}$ (or $2.16-3.16\,\rm{GeV}$). We also obtain masses and pole resides of the $sq\bar{s}\bar{q}$ and $qq\bar{q}\bar{q}$ vector tetraquark states and hope those fully-light vector tetraquark states can be observed in the future.

\section*{Appendix}
The analytical expressions of the QCD spectral densities for the P-wave fully-light vector tetraquark states in the case of the partial derivatives,
\begin{eqnarray}
\rho^1_{ss\bar{s}\bar{s}}(s)&=& \frac{31s^5}{806400\pi^6}+\frac{s^3\,m_s\langle\bar{s}s\rangle}{40\pi^4}
+\frac{s^3}{768\pi^4}\langle\frac{\alpha_{s}GG}{\pi}\rangle
-\frac{25s^2\,m_s\langle\bar{s}g_{s}\sigma Gs\rangle}{576\pi^4}
 \nonumber\\
&&+\frac{s m_s\langle\bar{s}s\rangle}{9\pi^2}\langle\frac{\alpha_{s}GG}{\pi}\rangle  +\frac{7s\langle\bar{s}s\rangle \langle\bar{s}g_{s}\sigma Gs\rangle}{9\pi^2} -\frac{7\langle\bar{s}s\rangle^2}{18}\langle\frac{\alpha_{s}GG}{\pi}\rangle \nonumber\\
&& -\frac{5\langle\bar{s}g_{s}\sigma Gs\rangle^2}{12\pi^2} -\frac{142m_s\langle\bar{s}s\rangle^2\langle\bar{s}g_{s}\sigma Gs\rangle}{27}\delta(s) \, ,
\end{eqnarray}

\begin{eqnarray}
\rho^2_{ss\bar{s}\bar{s}}(s)&=& \frac{s^5}{23040\pi^6}+\frac{s^3\,m_s\langle\bar{s}s\rangle}{24\pi^4}
+\frac{49s^3}{92160\pi^4}\langle\frac{\alpha_{s}GG}{\pi}\rangle
-\frac{109s^2\,m_s\langle\bar{s}g_{s}\sigma Gs\rangle}{1152\pi^4} \nonumber\\
&&-\frac{s^2\langle\bar{s}s\rangle^2}{9\pi^2}
+\frac{s m_s\langle\bar{s}s\rangle}{1728\pi^2}\langle\frac{\alpha_{s}GG}{\pi}\rangle  +\frac{17s\langle\bar{s}s\rangle \langle\bar{s}g_{s}\sigma Gs\rangle}{18\pi^2} -\frac{\langle\bar{s}s\rangle^2}{9}\langle\frac{\alpha_{s}GG}{\pi}\rangle \nonumber\\
&& -\frac{13\langle\bar{s}g_{s}\sigma Gs\rangle^2}{48\pi^2} -\frac{25m_s\langle\bar{s}s\rangle^2\langle\bar{s}g_{s}\sigma Gs\rangle}{9}\delta(s) \, ,
\end{eqnarray}

\begin{eqnarray}
\rho^3_{ss\bar{s}\bar{s}}(s)&=& \frac{53s^5}{201600\pi^6}+\frac{3s^3\,m_s\langle\bar{s}s\rangle}{10\pi^4}
+\frac{s^3}{192\pi^4}\langle\frac{\alpha_{s}GG}{\pi}\rangle
-\frac{35s^2\,m_s\langle\bar{s}g_{s}\sigma Gs\rangle}{288\pi^4} \nonumber\\
&&-\frac{s m_s\langle\bar{s}s\rangle}{54\pi^2}\langle\frac{\alpha_{s}GG}{\pi}\rangle -\frac{128m_s\langle\bar{s}s\rangle^2\langle\bar{s}g_{s}\sigma Gs\rangle}{3}\delta(s) \, , \end{eqnarray}
\begin{eqnarray}
\rho^4_{ss\bar{s}\bar{s}}(s)&=& \frac{s^5}{20160\pi^6}+\frac{3s^3\,m_s\langle\bar{s}s\rangle}{40\pi^4}
+\frac{s^3}{23040\pi^4}\langle\frac{\alpha_{s}GG}{\pi}\rangle
-\frac{5s^2\,m_s\langle\bar{s}g_{s}\sigma Gs\rangle}{128\pi^4} \nonumber\\
&&+\frac{19s m_s\langle\bar{s}s\rangle}{864\pi^2}\langle\frac{\alpha_{s}GG}{\pi}\rangle  +\frac{s\langle\bar{s}s\rangle \langle\bar{s}g_{s}\sigma Gs\rangle}{3\pi^2} -\frac{\langle\bar{s}s\rangle^2}{2}\langle\frac{\alpha_{s}GG}{\pi}\rangle \nonumber\\
&& -\frac{\langle\bar{s}g_{s}\sigma Gs\rangle^2}{2\pi^2} -\frac{146m_s\langle\bar{s}s\rangle^2\langle\bar{s}g_{s}\sigma Gs\rangle}{9}\delta(s) \, ,
\end{eqnarray}

\begin{eqnarray}
\rho^1_{qs\bar{q}\bar{s}}(s)&=& \frac{31s^5}{3225600\pi^6}+\frac{s^3\,m_s(5\langle\bar{s}s\rangle-2\langle\bar{q}q\rangle)}{960\pi^4}
+\frac{s^3}{3072\pi^4}\langle\frac{\alpha_{s}GG}{\pi}\rangle \nonumber\\
&&-\frac{s^2\,m_s(18\langle\bar{q}g_{s}\sigma Gq\rangle+7\langle\bar{s}g_{s}\sigma Gs\rangle)}{4608\pi^4}+\frac{s\,m_s(13\langle\bar{q}q\rangle+11\langle\bar{s}s\rangle)}{1728\pi^2}\langle\frac{\alpha_{s}GG}{\pi}\rangle  \nonumber\\
&&+\frac{7s(\langle\bar{s}s\rangle \langle\bar{q}g_{s}\sigma G q\rangle +\langle\bar{q}q\rangle \langle\bar{s}g_{s}\sigma Gs\rangle)}{72\pi^2}-\frac{7\langle\bar{q}q\rangle\langle\bar{s}s\rangle}{72}\langle\frac{\alpha_{s}GG}{\pi}\rangle -\frac{5\langle\bar{q}g_{s}\sigma Gq\rangle\langle\bar{s}g_{s}\sigma Gs\rangle}{48\pi^2} \nonumber\\
&&-\frac{m_s(48\langle\bar{q}q\rangle^2\langle\bar{s}g_{s}\sigma Gs\rangle-12\langle\bar{s}s\rangle^2\langle\bar{q}g_{s}\sigma Gq\rangle+48\langle\bar{q}q\rangle\langle\bar{s}s\rangle\langle\bar{q}g_{s}\sigma Gq\rangle-13\langle\bar{q}q\rangle\langle\bar{s}s\rangle\langle\bar{s}g_{s}\sigma Gs\rangle)}{108}\delta(s) \, , \nonumber\\
\end{eqnarray}

\begin{eqnarray}
\rho^2_{qs\bar{q}\bar{s}}(s)&=& \frac{s^5}{92160\pi^6}+\frac{s^3\,m_s(3\langle\bar{s}s\rangle-\langle\bar{q}q\rangle)}{384\pi^4}
+\frac{49s^3}{368640\pi^4}\langle\frac{\alpha_{s}GG}{\pi}\rangle
 \nonumber\\
&&-\frac{s^2\,m_s(41\langle\bar{q}g_{q}\sigma Gs\rangle+68\langle\bar{s}g_{s}\sigma Gs\rangle)}{9216\pi^4}-\frac{s^2\langle\bar{q}q\rangle\langle\bar{s}s\rangle}{36\pi^2}
+\frac{s m_s(10\langle\bar{q}q\rangle-9\langle\bar{s}s\rangle)}{13824\pi^2}\langle\frac{\alpha_{s}GG}{\pi}\rangle \nonumber\\
&&+\frac{17s(\langle\bar{s}s\rangle \langle\bar{q}g_{s}\sigma Gq\rangle+\langle\bar{q}q\rangle \langle\bar{s}g_{s}\sigma Gs\rangle)}{144\pi^2} -\frac{\langle\bar{q}q\rangle\langle\bar{s}s\rangle}{36}\langle\frac{\alpha_{s}GG}{\pi}\rangle  -\frac{13\langle\bar{q}g_{s}\sigma Gq\rangle\langle\bar{s}g_{s}\sigma Gs\rangle}{192\pi^2} \nonumber\\
&& -\frac{m_s(24\langle\bar{q}q\rangle^2\langle\bar{s}g_{s}\sigma Gs\rangle-10\langle\bar{s}s\rangle^2\langle\bar{q}g_{s}\sigma Gq\rangle+24\langle\bar{q}q\rangle\langle\bar{s}s\rangle\langle\bar{q}g_{s}\sigma Gq\rangle-13\langle\bar{q}q\rangle\langle\bar{s}s\rangle\langle\bar{s}g_{s}\sigma Gs\rangle)}{72}\delta(s) \, , \nonumber\\
\end{eqnarray}

\begin{eqnarray}
\rho^3_{qs\bar{q}\bar{s}}(s)&=& \frac{53s^5}{806400\pi^6}+\frac{3s^3\,m_s\langle\bar{s}s\rangle}{80\pi^4}
+\frac{s^3}{768\pi^4}\langle\frac{\alpha_{s}GG}{\pi}\rangle
-\frac{35s^2\,m_s\langle\bar{s}g_{s}\sigma Gs\rangle}{2304\pi^4} \nonumber\\
&&-\frac{s m_s\langle\bar{s}s\rangle}{432\pi^2}\langle\frac{\alpha_{s}GG}{\pi}\rangle -\frac{8m_s(\langle\bar{q}q\rangle^2\langle\bar{s}g_{s}\sigma Gs\rangle+\langle\bar{q}q\rangle\langle\bar{s}s\rangle\langle\bar{q}g_{s}\sigma Gq\rangle)}{3}\delta(s) \, ,
\end{eqnarray}

\begin{eqnarray}
\rho^4_{qs\bar{q}\bar{s}}(s)&=& \frac{s^5}{80640\pi^6}+\frac{s^3\,m_s(2\langle\bar{q}q\rangle+7\langle\bar{s}s\rangle)}{960\pi^4}
+\frac{s^3}{92160\pi^4}\langle\frac{\alpha_{s}GG}{\pi}\rangle
-\frac{5s^2\,m_s\langle\bar{s}g_{s}\sigma Gs\rangle}{1024\pi^4} \nonumber\\
&&+\frac{s m_s(11\langle\bar{q}q\rangle+27\langle\bar{s}s\rangle)}{13824\pi^2}\langle\frac{\alpha_{s}GG}{\pi}\rangle  +\frac{s(\langle\bar{s}s\rangle \langle\bar{q}g_{s}\sigma Gq\rangle+ \langle\bar{q}q\rangle\langle\bar{s}g_{s}\sigma Gs\rangle)}{24\pi^2} \nonumber\\
&&-\frac{\langle\bar{q}q\rangle\langle\bar{s}s\rangle}{8}\langle\frac{\alpha_{s}GG}{\pi}\rangle  -\frac{\langle\bar{q}g_{s}\sigma Gq\rangle\langle\bar{s}g_{s}\sigma Gs\rangle}{8\pi^2}\nonumber\\
&& -\frac{m_s(192\langle\bar{q}q\rangle^2\langle\bar{s}g_{s}\sigma Gs\rangle+22\langle\bar{s}s\rangle^2\langle\bar{q}g_{s}\sigma Gq\rangle+192\langle\bar{q}q\rangle\langle\bar{s}s\rangle\langle\bar{q}g_{s}\sigma Gq\rangle+32\langle\bar{q}q\rangle\langle\bar{s}s\rangle\langle\bar{s}g_{s}\sigma Gs\rangle)}{216}\delta(s) \, ,\nonumber\\
\end{eqnarray}

\begin{eqnarray}
\rho^5_{qs\bar{q}\bar{s}}(s)&=& \frac{s^5}{691600\pi^6}+\frac{s^3\,m_s(2\langle\bar{q}q\rangle+\langle\bar{s}s\rangle)}{960\pi^4}
+\frac{s^3}{46080\pi^4}\langle\frac{\alpha_{s}GG}{\pi}\rangle
-\frac{7s^2\,m_s\langle\bar{s}g_{s}\sigma Gs\rangle}{18432\pi^4} \nonumber\\
&&+\frac{s m_s(8\langle\bar{q}q\rangle+3\langle\bar{s}s\rangle)}{6912\pi^2}\langle\frac{\alpha_{s}GG}{\pi}\rangle  +\frac{s(\langle\bar{s}s\rangle \langle\bar{q}g_{s}\sigma Gq\rangle+ \langle\bar{q}q\rangle\langle\bar{s}g_{s}\sigma Gs\rangle)}{24\pi^2} \nonumber\\
&&-\frac{29\langle\bar{q}q\rangle\langle\bar{s}s\rangle}{576}\langle\frac{\alpha_{s}GG}{\pi}\rangle  -\frac{\langle\bar{q}g_{s}\sigma Gq\rangle\langle\bar{s}g_{s}\sigma Gs\rangle}{16\pi^2}\nonumber\\
&& -\frac{m_s(24\langle\bar{q}q\rangle^2\langle\bar{s}g_{s}\sigma Gs\rangle-12\langle\bar{s}s\rangle^2\langle\bar{q}g_{s}\sigma Gq\rangle+24\langle\bar{q}q\rangle\langle\bar{s}s\rangle\langle\bar{q}g_{s}\sigma Gq\rangle-13\langle\bar{q}q\rangle\langle\bar{s}s\rangle\langle\bar{s}g_{s}\sigma Gs\rangle)}{216}\delta(s) \, ,\nonumber\\
\end{eqnarray}

\begin{eqnarray}
\rho^6_{qs\bar{q}\bar{s}}(s)&=& \frac{19s^5}{38707200\pi^6}+\frac{s^3\,m_s(9\langle\bar{s}s\rangle-5\langle\bar{q}q\rangle)}{11520\pi^4}
+\frac{7s^3}{221184\pi^4}\langle\frac{\alpha_{s}GG}{\pi}\rangle
-\frac{s^2\,m_s(467\langle\bar{q}g_{s}\sigma Gq\rangle-230\langle\bar{s}g_{s}\sigma Gs\rangle)}{110592\pi^4} \nonumber\\
&&+\frac{5s^2\langle\bar{q}q\rangle\langle\bar{s}s\rangle}{216\pi^2}-\frac{s m_s(227\langle\bar{q}q\rangle-20\langle\bar{s}s\rangle)}{41472\pi^2}\langle\frac{\alpha_{s}GG}{\pi}\rangle \nonumber\\
&&-\frac{s(\langle\bar{q}q\rangle \langle\bar{q}g_{s}\sigma Gq\rangle+\langle\bar{s}s\rangle \langle\bar{s}g_{s}\sigma Gs\rangle+353\langle\bar{s}s\rangle \langle\bar{q}g_{s}\sigma Gq\rangle+ 353\langle\bar{q}q\rangle\langle\bar{s}g_{s}\sigma Gs\rangle)}{5184\pi^2} \nonumber\\
&&+\frac{\langle\bar{q}q\rangle\langle\bar{s}s\rangle}{24}\langle\frac{\alpha_{s}GG}{\pi}\rangle  -\frac{(\langle\bar{q}g_{s}\sigma Gq\rangle^2+\langle\bar{s}g_{s}\sigma Gs\rangle^2-82\langle\bar{q}g_{s}\sigma Gq\rangle\langle\bar{s}g_{s}\sigma Gs\rangle)}{1536\pi^2}\nonumber\\
&& +\frac{m_s(48\langle\bar{q}q\rangle^2\langle\bar{s}g_{s}\sigma Gs\rangle-17\langle\bar{s}s\rangle^2\langle\bar{q}g_{s}\sigma Gq\rangle+47\langle\bar{q}q\rangle\langle\bar{s}s\rangle\langle\bar{q}g_{s}\sigma Gq\rangle-25\langle\bar{q}q\rangle\langle\bar{s}s\rangle\langle\bar{s}g_{s}\sigma Gs\rangle)}{216}\delta(s)\, ,\nonumber\\
\end{eqnarray}

\begin{eqnarray}
\rho^1_{qq\bar{q}\bar{q}}(s)&=& \frac{31s^5}{806400\pi^6}
+\frac{s^3}{768\pi^4}\langle\frac{\alpha_{s}GG}{\pi}\rangle+\frac{7s\langle\bar{q}q\rangle \langle\bar{q}g_{s}\sigma Gq\rangle}{9\pi^2}
-\frac{7\langle\bar{q}q\rangle^2}{18}\langle\frac{\alpha_{s}GG}{\pi}\rangle -\frac{5\langle\bar{q}g_{s}\sigma Gq\rangle^2}{12\pi^2} \, ,
\end{eqnarray}

\begin{eqnarray}
\rho^2_{qq\bar{q}\bar{q}}(s)&=& \frac{s^5}{23040\pi^6}
+\frac{49s^3}{92160\pi^4}\langle\frac{\alpha_{s}GG}{\pi}\rangle-\frac{s^2\langle\bar{q}q\rangle^2}{9\pi^2}
 +\frac{17s\langle\bar{q}q\rangle \langle\bar{q}g_{s}\sigma Gq\rangle}{18\pi^2} \nonumber\\
&&-\frac{\langle\bar{q}q\rangle^2}{9}\langle\frac{\alpha_{s}GG}{\pi}\rangle  -\frac{13\langle\bar{q}g_{s}\sigma Gq\rangle^2}{48\pi^2} \, ,
\end{eqnarray}

\begin{eqnarray}
\rho^3_{qq\bar{q}\bar{q}}(s)&=& \frac{53s^5}{201600\pi^6}
+\frac{s^3}{192\pi^4}\langle\frac{\alpha_{s}GG}{\pi}\rangle \, ,
\end{eqnarray}

\begin{eqnarray}
\rho^4_{qq\bar{q}\bar{q}}(s)&=& \frac{s^5}{20160\pi^6}
+\frac{s^3}{23040\pi^4}\langle\frac{\alpha_{s}GG}{\pi}\rangle+\frac{s\langle\bar{q}q\rangle \langle\bar{q}g_{s}\sigma Gq\rangle}{3\pi^2} -\frac{\langle\bar{q}q\rangle^2}{2}\langle\frac{\alpha_{s}GG}{\pi}\rangle -\frac{\langle\bar{q}g_{s}\sigma Gq\rangle^2}{2\pi^2} \, . \nonumber\\
\end{eqnarray}

With the simple replacements,
\begin{eqnarray}
\rho^{1/2/3/4}_{ss\bar{s}\bar{s}}(s)&\to&\rho^{1/2/3/4}_{ss\bar{s}\bar{s}}(s)+\tilde{\rho}^{1/2/3/4}_{ss\bar{s}\bar{s}}(s)\, , \nonumber \\
\rho^{1/2/3/4/5/6}_{qs\bar{q}\bar{s}}(s)&\to&\rho^{1/2/3/4/5/6}_{qs\bar{q}\bar{s}}(s)+\tilde{\rho}^{1/2/3/4/5/6}_{qs\bar{q}\bar{s}}(s)\, , \nonumber \\
\rho^{1/2/3/4}_{qq\bar{q}\bar{q}}(s)&\to&\rho^{1/2/3/4}_{qq\bar{q}\bar{q}}(s)+\tilde{\rho}^{1/2/3/4}_{qq\bar{q}\bar{q}}(s)\, ,
\end{eqnarray}
we obtain the corresponding QCD spectral densities for the currents with the covariant derivatives, where the additional terms,

\begin{eqnarray}
\tilde{\rho}^1_{ss\bar{s}\bar{s}}(s)&=&0\, ,
\end{eqnarray}

\begin{eqnarray}
\tilde{\rho}^2_{ss\bar{s}\bar{s}}(s)&=&-\frac{77s^3}{184320\pi^4}\langle\frac{\alpha_{s}GG}{\pi}\rangle
+\frac{253 s m_s\langle\bar{s}s\rangle}{6912\pi^2}\langle\frac{\alpha_{s}GG}{\pi}\rangle -\frac{11\langle\bar{s}s\rangle^2}{96}\langle\frac{\alpha_{s}GG}{\pi}\rangle
\, ,
\end{eqnarray}

\begin{eqnarray}
\tilde{\rho}^3_{ss\bar{s}\bar{s}}(s)&=&0\, ,
\end{eqnarray}

\begin{eqnarray}
\tilde{\rho}^4_{ss\bar{s}\bar{s}}(s)&=&-\frac{51s^3}{61440\pi^4}\langle\frac{\alpha_{s}GG}{\pi}\rangle
+\frac{7s^2\,m_s\langle\bar{s}g_{s}\sigma Gs\rangle}{2304\pi^4} +\frac{719s m_s\langle\bar{s}s\rangle}{6912\pi^2}\langle\frac{\alpha_{s}GG}{\pi}\rangle \nonumber\\
&& +\frac{5s\langle\bar{s}s\rangle \langle\bar{s}g_{s}\sigma Gs\rangle}{864\pi^2} -\frac{\langle\bar{s}g_{s}\sigma Gs\rangle^2}{384\pi^2} -\frac{m_s\langle\bar{s}s\rangle^2\langle\bar{s}g_{s}\sigma Gs\rangle}{108}\delta(s) \, ,
\end{eqnarray}

\begin{eqnarray}
\tilde{\rho}^1_{qs\bar{q}\bar{s}}(s)&=&0\, ,
\end{eqnarray}

\begin{eqnarray}
\tilde{\rho}^2_{qs\bar{q}\bar{s}}(s)&=&-\frac{77s^3}{737280\pi^4}\langle\frac{\alpha_{s}GG}{\pi}\rangle
+\frac{77 s m_s(2\langle\bar{q}q\rangle-\langle\bar{s}s\rangle)}{13824\pi^2}\langle\frac{\alpha_{s}GG}{\pi}\rangle -\frac{11\langle\bar{s}s\rangle^2}{384}\langle\frac{\alpha_{s}GG}{\pi}\rangle
\, ,
\end{eqnarray}

\begin{eqnarray}
\tilde{\rho}^3_{qs\bar{q}\bar{s}}(s)&=&0\, ,
\end{eqnarray}

\begin{eqnarray}
\tilde{\rho}^4_{qs\bar{q}\bar{s}}(s)&=&-\frac{51s^3}{245760\pi^4}\langle\frac{\alpha_{s}GG}{\pi}\rangle
 +\frac{sm_s(\langle\bar{q}q\rangle-153\langle\bar{s}s\rangle)}{13824\pi^2}\langle\frac{\alpha_{s}GG}{\pi}\rangle
 \nonumber\\
&& +\frac{5s\langle\bar{q}q\rangle \langle\bar{s}g_{s}\sigma Gs\rangle}{3456\pi^2} -\frac{\langle\bar{q}g_{q}\sigma Gs\rangle\langle\bar{s}g_{s}\sigma Gs\rangle}{1536\pi^2} -\frac{m_s\langle\bar{q}q\rangle \langle\bar{s}s\rangle \langle\bar{s}g_{s}\sigma Gs\rangle}{864}\delta(s) \, ,
\end{eqnarray}

\begin{eqnarray}
\tilde{\rho}^5_{qs\bar{q}\bar{s}}(s)&=&0\, ,
\end{eqnarray}

\begin{eqnarray}
\tilde{\rho}^6_{qs\bar{q}\bar{s}}(s)&=&\frac{7s^3}{2211840\pi^4}\langle\frac{\alpha_{s}GG}{\pi}\rangle
+\frac{7s^2\,m_s \langle\bar{q}g_{s}\sigma Gq\rangle}{36864\pi^4}-\frac{sm_s(7\langle\bar{q}q\rangle-4\langle\bar{s}s\rangle)}{20736\pi^2}\langle\frac{\alpha_{s}GG}{\pi}\rangle \nonumber\\
&&-\frac{(5\langle\bar{s}s\rangle \langle\bar{q}g_{s}\sigma Gq\rangle+ 5\langle\bar{q}q\rangle\langle\bar{s}g_{s}\sigma Gs\rangle)}{3456\pi^2}+\frac{7\langle\bar{q}q\rangle\langle\bar{s}s\rangle}{6912}\langle\frac{\alpha_{s}GG}{\pi}\rangle +\frac{\langle\bar{q}g_{s}\sigma Gq\rangle\langle\bar{s}g_{s}\sigma Gs\rangle}{768\pi^2}\nonumber\\
&& +\frac{m_s(2\langle\bar{q}q\rangle^2\langle\bar{s}g_{s}\sigma Gs\rangle-\langle\bar{s}s\rangle^2\langle\bar{q}g_{s}\sigma Gq\rangle+2\langle\bar{q}q\rangle\langle\bar{s}s\rangle\langle\bar{q}g_{s}\sigma Gq\rangle-\langle\bar{q}q\rangle\langle\bar{s}s\rangle\langle\bar{s}g_{s}\sigma Gs\rangle)}{2592}\delta(s)\, , \nonumber\\
\end{eqnarray}

\begin{eqnarray}
\tilde{\rho}^1_{qq\bar{q}\bar{q}}(s)&=&0\, ,
\end{eqnarray}

\begin{eqnarray}
\tilde{\rho}^2_{qq\bar{q}\bar{q}}(s)&=&-\frac{77s^3}{184320\pi^4}\langle\frac{\alpha_{s}GG}{\pi}\rangle -\frac{11\langle\bar{q}q\rangle^2}{96}\langle\frac{\alpha_{s}GG}{\pi}\rangle
\, ,
\end{eqnarray}

\begin{eqnarray}
\tilde{\rho}^3_{qq\bar{q}\bar{q}}(s)&=&0\, ,
\end{eqnarray}

\begin{eqnarray}
\tilde{\rho}^4_{qq\bar{q}\bar{q}}(s)&=&-\frac{51s^3}{61440\pi^4}\langle\frac{\alpha_{s}GG}{\pi}\rangle
+\frac{5s\langle\bar{q}q\rangle \langle\bar{q}g_{s}\sigma Gq\rangle}{864\pi^2} -\frac{\langle\bar{q}g_{s}\sigma Gq\rangle^2}{384\pi^2} \, .
\end{eqnarray}

\section*{Acknowledgements}
This  work is supported by National Natural Science Foundation, Grant Number 12175068, and Postgraduate Students Innovative Capacity Foundation of Hebei Education Department, Grant
Number CXZZBS2023146.


\begin{thebibliography}{99}


\bibitem{BabarY2175-2006} B. Aubert, et al, Phys. Rev. {\bf D74} (2006) 091103.

\bibitem{ssss-ZGWANG-2006} Z. G. Wang, Nucl. Phys. {\bf A791} (2007) 106.

\bibitem{Review-CPC} M. Ablikim, et al, Chin. Phys. {\bf C44} (2020)  040001.

\bibitem{BESIIY2175-2007} M. Ablikim, et al, Phys. Rev. Lett. {\bf100} (2008) 102003.

\bibitem{BelleY2175-2008} C. P. Shen, et al, Phys. Rev. {\bf D80} (2009) 031101.

\bibitem{Y2400-2008-combine} C. P. Shen and C. Z. Yuan, Chin. Phys. {\bf C34} (2010) 1045.

\bibitem{BESIIIY2175(2239)-2018}  M. Ablikim, et al, Phys. Rev. {\bf D99} (2019) 032001.

\bibitem{BESIII2232-2019}  M. Ablikim, et al, Phys. Rev. {\bf D100} (2019) 032009.

\bibitem{PDG-2020} P. A. Zyla et al, Prog. Theor. Exp. Phys. {\bf2020} (2020) 083C01.

\bibitem{BABAR2232-2020} J. P. Lees, et al, Phys. Rev. {\bf D101} (2020) 012011.

\bibitem{BESIII2356-2022} M. Ablikim, et al, Phys. Rev. {\bf D107} (2023) 112001.


\bibitem{BES-1835} M. Ablikim, et al, Phys. Rev. Lett. {\bf 95} (2005) 262001.

\bibitem{BESIIIX1835/2120/2370-2010} M. Ablikim, et al, Phys. Rev. Lett. {\bf106} (2011) 072002.

\bibitem{BESIIIX2100/2500etc-2016} M. Ablikim, et al, Phys. Rev. {\bf D93} (2016) 112011.


\bibitem{Y2175CHX2008} H. X. Chen, X. Liu, A. Hosaka and S. L. Zhu, Phys. Rev. {\bf D78} (2008) 034012.

\bibitem{Y2175CHX2018} H. X. Chen, C. P. Shen and S. L. Zhu, Phys. Rev.  {\bf D98} (2018) 014011.

\bibitem{Y2175CHX2022} N. Su and H. X. Chen, Phys. Rev. {\bf D106} (2022)  014023.


\bibitem{X2100/2239-AZIZI-2019} K. Azizi, S. S. Agaev and H. Sundu, Nucl. Phys. {\bf B948} (2019) 114789.

\bibitem{Y2175-AZIZI-2019} S. S. Agaev, K. Azizi and H. Sundu, Phys. Rev. {\bf D101} (2020) 074012.

\bibitem{ssss-LQF-2019} Q. F. Lu, K. L. Wang and Y. B. Dong, Chin. Phys. {\bf C44} (2020) 024101.

\bibitem{2170-LXQ-2018} H. W. Ke and X. Q. Li, Phys. Rev. {\bf D99} (2019) 036014.

\bibitem{2175-1536-PJL-2010} C. R. Deng, J. L. Ping, F. Wang and T. Goldman, Phys. Rev. {\bf D82} (2010) 074001.

\bibitem{2175-PJL-2013} C. R. Deng, J. L. Ping, Y. Yang and F. Wang, Phys. Rev. {\bf D88} (2013) 074007.

\bibitem{2170-LQF-2017} Y. B. Dong, A. Faessler, T. Gutsche, Q. F. Lu and V. E. Lyubovitskij, Phys. Rev. {\bf D96} (2017) 074027.

\bibitem{2175-DGJ-2007-1} G. J. Ding and M. L. Yan, Phys. Lett. {\bf B657} (2007) 49.

\bibitem{phi-CQP-2019} C. Q. Pang, Phys. Rev. {\bf D99} (2019) 074015.

\bibitem{phiKK-2008} A. M. Torres, K. P. Khemchandani, L. S. Geng, M. Napsuciale and E. Oset,
Phys. Rev. {\bf D78} (2008) 074031.

\bibitem{phiKK-2009} S. Gomez-Avila, M. Napsuciale and E. Oset, Phys. Rev. {\bf D79} (2009) 034018.

\bibitem{2175-DGJ-2007} G. J. Ding and M. L. Yan, Phys. Lett. {\bf B650} (2007) 390.

\bibitem{ssss-ZGWANG-2019} Z. G. Wang, Adv. High Energy Phys. {\bf2020} (2020) 6438730.

\bibitem{WZG-CTP-SA} Z. G. Wang, Commun. Theor. Phys. {\bf 59} (2013) 451.


\bibitem{Color-Spin} A. De Rujula, H. Georgi and S. L. Glashow, Phys. Rev. {\bf D12} (1975) 147.

\bibitem{Jaffe1977-1} R. L. Jaffe, Phys. Rev. {\bf D15} (1977) 267.

\bibitem{Jaffe1977-2} R. L. Jaffe, Phys. Rev. {\bf D15} (1977) 281.

\bibitem{Y4260-ZGWANG-2018} Z. G. Wang, Eur. Phys. J. {\bf C78} (2018) 518.

\bibitem{WZG-CTP-4100}  Z. G. Wang, Commun. Theor. Phys. {\bf 71} (2019) 1319.

\bibitem{Y4260P-wave-ZGWANG-2018} Z. G. Wang, Eur. Phys. J. {\bf C78} (2018) 933.

\bibitem{Y4220P-wave-ZGWANG-2018} Z. G. Wang, Eur. Phys. J. {\bf C79} (2019) 29.

\bibitem{NPB-ZGWANG-2021} Z. G. Wang, Nucl. Phys. {\bf B973} (2021) 115592.


\bibitem{WZG-Dwave-baryon} Z. G. Wang, F. Lu and Y. Liu, Eur. Phys. J. {\bf C83} (2023) 689.

\bibitem{XQWZG-Pwave-Bbaryon} Q. Xin, Z. G. Wang and F. Lu, Chin. Phys. {\bf C47} (2023) 093106.

\bibitem{QCDSR-SVZ79} M. A. Shifman, A. I. Vainshtein and V. I. Zakharov, Nucl. Phys. {\bf B147} (1979) 385;
Nucl. Phys. {\bf B147} (1979) 448.

\bibitem{QCDSR-Reinders85} L. J. Reinders, H. Rubinstein and S. Yazaki, Phys. Rept. {\bf 127} (1985) 1.

\bibitem{Pascual-1984} P. Pascual and R. Tarrach, ``QCD: Renormalization for the practitioner", Springer Berlin Heidelberg (1984).

\bibitem{WangHuang3900}  Z. G. Wang and T. Huang,  Phys. Rev. {\bf D89} (2014)  054019.

\bibitem{Wang-tetraquark-QCDSR-2} Z. G. Wang, Eur. Phys. J. {\bf C74} (2014)  2874.

\bibitem{WZG-CPC-Zcs} Z. G. Wang, Chin. Phys. {\bf C45} (2021)  073107.

\bibitem{WXW-EPJA} X. W. Wang, Z. G. Wang and G. L. Yu, Eur. Phys. J. {\bf A57} (2021)  275.


\bibitem{QCDSR-Colangelo-Review} P. Colangelo and A. Khodjamirian, hep-ph/0010175.


\bibitem{WZG-EPJC-Scalar} Z. G. Wang, Eur. Phys. J. {\bf C76} (2016) 427.

\bibitem{WZG-Vector-CPL}  Z. G. Wang and S. L. Wan, Chin. Phys. Lett. {\bf 23} (2006) 3208.





\end{thebibliography}
\end{document}